\documentclass[12pt]{article}
\usepackage{amsmath,amssymb,amsthm}
\usepackage{booktabs}
\usepackage{multirow}
\usepackage{siunitx}
\usepackage{graphicx}
\usepackage{geometry}
\usepackage{algorithm}
\usepackage{algorithmicx}
\usepackage{comment}
\usepackage{subcaption}
\usepackage{algpseudocode}
\usepackage{bm}
\usepackage[authoryear]{natbib}
\usepackage{hyperref}
\usepackage{adjustbox}
\usepackage{float}
\usepackage{booktabs}
\usepackage{longtable}
\usepackage{pdflscape}
\usepackage{epstopdf}

\usepackage{threeparttable}

\title{Interval-valued circular time series: A toroidal score-driven approach with an application to wind direction}

\author{
Thasmika Mohan$^{1}$, Najmeh Nakhaei Rad$^{1}\footnote{Correspoding Author: najmeh.nakhaeirad@up.ac.za}$, 
Sphiwe Skhosana$^{1}$ \\
Yaser Samadi $^{2}$\\[4pt]
\small $^{1}$Department of Statistics, University of Pretoria, Pretoria 0002, South Africa. \\
\small $^{2}$School of Mathematical and Statistical Sciences, Southern Illinois University, Carbondale, IL, USA.
}

\date{}
\begin{document}
\maketitle  \begin{abstract} Circular data recorded over time, such as wind direction, are increasingly observed in intervals rather than as single point-valued observations, yet no existing framework accommodates data that are simultaneously circular, interval-valued, and
temporally dependent. Time series models for point-valued circular data,
including recently proposed score-driven approaches, assume that each
observation is a single angle. This paper addresses this gap by proposing a dynamic conditional score framework for interval-valued circular time series. The lower
and upper boundary sequences of the observed intervals are treated as
circular time series governed by score-driven dynamics, which preserves the
periodicity of the circular domain and the dependency of boundaries,  in which the boundaries are
coupled through a toroidal distribution and a jointly derived score.
Maximum likelihood estimation is discussed, and their
finite-sample performance is evaluated through simulation. An application to
interval-valued wind direction data illustrates the practical value of
retaining the interval structure 
within a fully rotationally invariant circular framework.
\end{abstract}

\section{Introduction}
\label{sec:introduction}

Circular variables, measured as angles or directions on the unit circle,
arise across a wide range of scientific disciplines, 
including wind direction \citep{nakhaei2022coming,nakhaei2026predicting}, animal movement \citep{ramkilawon2023sine,gascoyne2024modelling}, ocean current \citep{mohan2024poly}, geological orientations \citep{lark2014modelling}, geographical locations \citep{jeong2017spherical}, protein structures \citep{arashi2021mobius,rad2025bayesian}, and engineering \citep{mardia2009directional}. Wind direction is
among the most commonly studied circular variables, with applications
ranging from wind farm resource assessment and turbine yaw control to the prediction of wildfire spread \citep{harvey2024modelling}. A defining feature of circular data is the periodicity of the observations. This means observations recorded just below $\pi$ and just above $-\pi$ are, in
reality, close together on the circle, even though they appear far apart
when treated as points on the real line. Statistical methods that ignore
this topology can produce severely distorted inference, particularly when observations are spread widely around the circle \citep{mardia2009directional}.

Data consisting of directions $\theta_t, \theta_{t+1}, \ldots, \theta_{t+p}$
recorded at successive time points are observed as circular time series data in many of these same disciplines: time series of wind direction assist in
climate change analysis, time series of wave and cloud direction assist with
disaster management in oceanography, and the direction of criminal activity over time has been modelled using circular time series methods
\citep{brunsdon2006using}. A literature review is discussed in further detail for
such point-valued circular time series, in
Section~\ref{sec:background}.

In many real-world settings, however, point-valued circular observations are not available, and the data are instead recorded as intervals. High-frequency
sensor networks, of the kind used at modern wind farms, routinely summarise
wind direction recorded over short intervals (for example, every ten
minutes) by reporting the minimum and maximum observed angle, rather than
every individual reading. The resulting interval-valued circular observations
retain more information than a simple average or midpoint, but they introduce
two features that existing circular time series models are not designed to
handle jointly: internal variability within each observed interval,
and dependence between intervals. 

Recently, developments in circular time series modelling have shown that dependence in point-valued circular data can be handled naturally
through score-driven dynamics. \citet{harvey2013} and \citet{creal2013}
introduced the score-driven, or Dynamic Conditional Score (DCS), approach, also referred to as the Generalised Autoregressive Score (GAS) class,
in which the time-varying location parameter of a distribution is updated
using the conditional score of that distribution, and
\citet{harvey2024modelling} extended it to the circular setting. The approach
has three advantages that are directly relevant to the problem considered
here. First, it is formulated directly on the circle, without transformation
to the real line, so that the circular structure of the data is preserved and
the fitted model does not depend on an arbitrary choice of reference
direction \citep{harvey2024modelling}. Second, because the score of a
circular density is itself a periodic function, the resulting filter is
invariant to how the data are wrapped onto the circle: adding or subtracting a multiple of $2\pi$ to an observation leaves the score, and hence the entire
filtered path, unchanged \citep{harvey2024modelling}. This removes the need
to estimate the unobserved integer multiples of $2\pi$ separating a wrapped
observation from its underlying real-line counterpart, and thereby avoids the computationally expensive missing-data problem that affects earlier
transformation-based approaches, such as linked autoregressive moving average
(LARMA) models \citep{fisher1994time}. Third, this property follows from the
periodicity of the score itself, rather than from any property specific to the
von~Mises distribution, so it is not tied to a particular circular density. The same
construction can be extended to accommodate non-stationary behaviour, and extends to regime
switching and to bivariate cylindrical distributions, combining a circular and
a linear variable \citep{harvey2023regime}. In this paper, we extend the DCS approach using a toroidal distribution, combining lower and upper bounds of intervals as two dependent circular variables. 

Separately, a parallel literature has developed for modelling interval-valued
data on the real line, through the earlier
set-valued approach of \citet{wang2016set} and interval-valued
autoregressive (IAR) framework of \citet{samadi2021analysis}, which
accommodates both the internal variation within each observed interval and
the serial dependence between intervals through a composite-likelihood
estimation procedure. This framework has since been extended to the circular
setting by \citet{mohan2026cinar}, who proposed a Circular Interval-Valued
Autoregressive (CINAR) model based on a link-function transformation of the
interval endpoints onto the real line. The CINAR model represents the interval as a single
transformed object at each time point, with dependence between the lower and
upper bounds captured implicitly through the joint transformation rather than modelled directly. However, this framework is not fully rotationally invariant on the circle \citet{harvey2024modelling}.
This creates a methodological gap. To fill this gap, this paper proposes a score-driven framework for
interval-valued circular time series, formulated directly on the circle
rather than through a transformation to the real line. Rather than
aggregating each observed interval to its midpoint or mean before modelling, the
proposed approach treats the lower and upper boundary sequences of the
observed intervals as two circular time series governed by score-driven
dynamics, allowing the fitted interval to evolve smoothly over time while
remaining consistent with the circular topology of the underlying variable.
The proposed framework makes the following contributions:
\begin{itemize}
  \item It extends the score-driven construction to interval-valued circular observations to develop a fully rotationally invariant circular time series model for interval-valued data.
  \item It develops a dependent model, in which the boundaries are coupled
        through a toroidal distribution of \cite{singh2002probabilistic} and hence develops maximum likelihood estimation for the proposed approach.
       
\end{itemize}
The finite-sample performance of the resulting estimators is evaluated
through a simulation study. Finally, the practical value of this approach is illustrated through two applications to interval-valued wind direction datasets.

The remainder of this paper is structured as follows.
Section~\ref{sec:background} reviews existing time series models for
point-valued circular data, including the score-driven
approach of \citet{harvey2024modelling} on which the proposed model is built,
and reviews existing approaches to interval-valued time series.
Section~\ref{sec:methodology} introduces the proposed
dependent dynamic score-driven models (IN-DCS) for interval-valued circular data and the maximum likelihood estimation procedure.
Section~\ref{sec:simulation} describes a simulation study assessing the
finite-sample performance of the proposed estimators.
Section~\ref{sec:application} applies the proposed models to two interval-valued
wind direction datasets. Finally, Section~\ref{sec:conclusion} provides a conclusion future research.

\section{Background}
\label{sec:background}

\subsection{Time series models for point-valued circular data}

Many time series models for point-valued circular observations have been
proposed in the literature. The earliest circular time series models were
introduced by \citet{fisher1992regression} and \citet{fisher1994time}, who proposed
three methods for modelling circular time series: the projected normal process, wrapping approach, and a one-to-one link function mapping approach. Subsequent developments have extended this line of work in several
directions. Within the Markov tradition, \citet{holzmann2006hidden} proposed a
hidden Markov approach for circular time series based on the stationary
Markov process of \citet{johnson1978some}; \citet{kato2008circular} introduced a
discrete-time Markov process using M\"obius transformations for modelling
wind direction; \citet{yeh2013drifting} introduced a circular Markov process in which
the transition probabilities belong to a class of generalised von~Mises
distributions; \citet{abe2017circular} introduced a Markov process whose transition
density is defined by two circular distributions; and \citet{ogata2023}
developed the first higher-order Markov process on the circle. In parallel,
\citet{artes2009autoregressive} extended the CAR($p$) model to incorporate covariates,
\citet{dimarzio2009} developed a non-parametric kernel-based approach to
circular time series estimation, and \citet{mazumder2016} derived a Bayesian
non-parametric model in which the latent states also constitute a circular
process. For multivariate circular data, \citet{jones2015class} considered
circular joint distributions, or ``circulas'', for analysing correlated
circular structures, and \citet{ogata2023} subsequently considered a
multivariate circular time series in which the cross-sectional and serial dependence are both expressed through circulas.

A more recent and, for the present paper, more directly relevant development
is the score-driven, or Dynamic Conditional Score (DCS), approach introduced
by \citet{harvey2013} and \citet{creal2013} and extended to the circular
setting by \citet{harvey2024modelling}. The score-driven approach is formulated directly on the circle, and it is
not restricted to data concentrated on a narrow arc, as the link-function approach  \citep{harvey2024modelling}.

For a
comprehensive review of circular time series, more broadly, see Chapter~7 of
\citet{fisher1994time}, Section~11.5.2 of \citet{jupp1980general}, Section~12.8 of
\citet{jammalamadaka2001topics}, and Section~10 of \citet{pewsey2021recent}.

\subsection{The score-driven model for point-valued circular data}
\label{sec:harvey-background}
The score-driven approach was introduced by \citet{harvey2013}
and \citet{creal2013}. A survey of the framework and its applications is
given by \citet{harvey2021review}. The framework was subsequently extended to circular observations. \citet{harveyhurnthiele2019} and \citet{harvey2024modelling} develop
score-driven models for circular time series, in which the
time-varying location parameter of a circular distribution is updated using the
conditional score of that distribution, so that the model is formulated
directly on the circle. Because the score of a circular density is itself a
periodic function, the resulting filter is invariant to how the observations
are wrapped onto the circle, which avoids the computationally expensive
missing-data problem affecting earlier transformation-based approaches.
\citet{harvey2023regime} extend the approach to regime-switching circular time
series, in which the intra-regime dynamics are themselves score-driven, and to
bivariate cylindrical distributions combining a circular and a linear variable;
an earlier working-paper version of the regime-switching model is given by
\citet{harvey2023regime}. These authors also consider the score-driven extension to include trend and seasonal
components.  \citet{caivano2015} develop robust score-driven
models with trend and seasonal components, in which the components may be
nonstationary.

Let $\theta_t \in [-\pi, \pi)$, $t = 1, \ldots, T$, denote a sequence of circular
observations, assumed to follow a von~Mises (vM) distribution with
probability density function
\begin{equation*}
  f(\theta_t; \mu, \upsilon) = \frac{1}{2\pi I_0(\upsilon)}
  \exp\{\upsilon \cos(\theta _t- \mu)\},
  \label{eq:vm-density}
\end{equation*}
where $I_k(\upsilon)$ denotes the modified Bessel function of order $k$, $\mu$
is the mean direction, and $\upsilon$ is a non-negative concentration
parameter. When $\upsilon = 0$ the distribution is uniform on the circle, and
as $\upsilon \to \infty$ the distribution approaches a Gaussian distribution
with mean $\mu$ and variance $1/\upsilon$.

The DCS model developed by \citet{harvey2024modelling} is outlined below. The observation equation is
\begin{equation*}
  \theta_t = \mu + \mu_{t|t-1} + \varepsilon_t, \qquad t = 1, \ldots, T,
  \label{eq:harvey-obs}
\end{equation*}
where $\mu\in [-\pi,\pi)$, $\varepsilon_t$ is a sequence of independent and identically distributed
circular random variables with location zero and the time-varying component
evolves according to
\begin{equation*}
  \mu_{t+1|t} = \phi\, \mu_{t|t-1} + \kappa\, u_t,
  \label{eq:harvey-recursion}
\end{equation*}
where $\kappa$ is the Kalman gain and $u_t \in [\pi, \pi)$ is the forcing variable and is proportional to the conditional score of the assumed circular
distribution for location. The time series coefficient is $\phi$ and for $|\phi|<1$, the model is stationary and thus $\mu_{t|t-1}$ is a linear combination of past observations with exponentially declining weights. For the von~Mises case, the score is obtained as 
\begin{equation*}
  u_t = \upsilon\sin\!\left(\theta_t - \mu - \mu_{t|t-1}\right).
  \label{eq:harvey-score}
\end{equation*}
Because the recursion is driven by the score rather than by a latent process,
the time-varying location is a deterministic function of past observations,
so the likelihood can be evaluated directly for any parameter value;
estimation therefore reduces to numerical optimisation of the likelihood, rather than to the high-dimensional integration required
by the projected normal approach or the multidimensional infinite sums
required by the wrapping approach \citep{fisher1994time}.

Since $\sin(\cdot)$ is a $2\pi$-periodic function, adding or
subtracting a multiple of $2\pi$ to an observation $\theta_t$ leaves $u_t$, and
hence the entire filtered path $\mu_{t|t-1}$, unchanged
\citep{harvey2024modelling}. This removes the need to estimate the unobserved
integer multiples of $2\pi$ separating a wrapped circular observation from
its underlying real-line counterpart. Using this property, the proposed interval-valued extension in Section~\ref{sec:methodology} is developed.

In the next section, we
propose a IN-DCS approach for the circular interval-valued time series data.
\section{Methodology}\label{sec:methodology}

In this section, we introduce a score-driven framework for interval-valued circular
time series. Rather than transforming each observed interval onto the real line, as
in the CINAR framework \cite{mohan2026cinar}, the proposed approach treats the lower and
upper boundary sequences of the observed intervals as circular time series,
thereby extending the score-driven method 
to interval-valued observations, that are circular and with dependency. Since the filter is built from the score of a toroidal distribution, the resulting recursions preserve the wrapping-invariance property established in Section~\ref{sec:harvey-background} hence the
observed boundary angles can be used directly.



Let $\Theta_t = [\theta_{Lt}, \theta_{Ut}]$, $t = 1, \dots, T$, denote a sequence of
interval-valued circular observations, where $\theta_{Lt}$ and $\theta_{Ut}$ denote
the lower and upper bounds of the interval recorded at time $t$, ordered so that
$\theta_{Lt} < \theta_{Ut}$ when both are measured anti-clockwise from a common
reference point on the circle. An interval crossing the circular cut, such as
$[350^\circ, 10^\circ]$, is represented as $[350^\circ, 370^\circ]$ to preserve
circular continuity, and the two angles are then wrapped onto $[-\pi, \pi)$ before
entering the filter. Rather than aggregating each interval to a single representative value, such as its
midpoint, we model the two boundary sequences $\{\theta_{Lt}\}_{t=1}^{T}$ and
$\{\theta_{Ut}\}_{t=1}^{T}$ jointly as circular time series governed by score-driven dynamics. This approach allows for dependency to be considered between the circular lower and upper bounds.

\subsection{Interval-valued dynamic score driven model}\label{sec:dependent}
To introduce the proposed IN-DCS model, we consider a suitable joint density for
$(e^L_{t}, e^{U}_{t})$. The bivariate Sine distribution of
\cite{singh2002probabilistic} extends the univariate von Mises density to a pair of dependent
circular random variables by adding a sine-product interaction term to the exponent:
\begin{equation}\label{eq:bivm}
  f_{sine}\!\left(e^L, e^U; \upsilon_L, \upsilon_U, \lambda\right)
  = \frac{1}{C(e^L, e^U, \lambda)}
    \exp\!\left\{\upsilon_L \cos(\varepsilon_L) + \upsilon_U \cos(e^U)
    + \lambda \sin(e^L) \sin(e^U)\right\},
\end{equation}
where $\upsilon_L > 0$ and $\upsilon_U > 0$ are the concentration parameters
of the lower and upper boundary error terms and $\lambda \in \mathbb{R}$ is a dependence
parameter. If $\lambda = 0$, the density results in two independent von Mises
distributions. The normalising constant is
\begin{equation}\label{eq:Z}
  C(\upsilon_L, \upsilon_U, \lambda)
  = 4\pi^2 \sum_{m=0}^{\infty} \binom{2m}{m}
    \left(\frac{\lambda^2}{4\upsilon_L \upsilon_U}\right)^{m}
    I_m(\upsilon_L) I_m(\upsilon_U),
\end{equation}
and we impose $\lambda^2 < \upsilon_L \upsilon_U$ to keep the joint density unimodal
at $(0, 0)$.

We retain the observation equations and filtering structure of
\cite{harvey2024modelling},
\begin{equation*}\label{eq:obs-dep}
  \theta_{bt} = \mu_b + \mu_{bt|t-1} + \varepsilon_{bt}, \qquad b \in \{L, U\},
\end{equation*}
\begin{equation*}\label{eq:filter-dep}
  \mu_{bt+1|t} = \phi_b \mu_{bt|t-1} + \kappa_b u_{bt}, \qquad \mu_{b1|0} = 0,
\end{equation*}
but now let $(\varepsilon_{Lt}, \varepsilon_{Ut})$ follow the joint density in
\eqref{eq:bivm} rather than two independent von Mises distributions. If the observations are outside the circular range $[-\pi, \pi)$, wrapping is applied. Since the score is continuous the filter is fully rotionally invariant on the circle. 

Assume $e_{bt} = \theta_{bt} - \mu_b - \mu_{bt|t-1}$, the conditional log-density at
time $t$ is
\begin{equation*}\label{eq:logdens}
   -\ln C(\upsilon_L, \upsilon_U, \lambda)
    + \upsilon_L \cos(e^{L}_{t}) + \upsilon_U \cos(e^{U}_{t})
    + \lambda \sin(e^{L}_t) \sin(e^{U}_{t}).
\end{equation*}

Differentiating with respect to the two time-varying locations gives the gradient
\begin{equation*}
\nabla_t=\begin{pmatrix}\upsilon_L\sin(e^L_t)-\lambda\cos(e^L_t)\sin(e^U_t)\\
\upsilon_U\sin(e^U_t)-\lambda\sin(e^L_t)\cos(e^U_t)\end{pmatrix}. 
\end{equation*}
and therefore,
\begin{align}
u^L_t&=\frac{\nabla^L_t}{\upsilon_L}=\sin(e^L_t)-\frac{\lambda}{\upsilon_L}\cos(e^L_t)\sin(e^U_t),\notag\\
u^U_t&=\frac{\nabla^U_t}{\upsilon_U}=\sin(e^U_t)-\frac{\lambda}{\upsilon_U}\sin(e^L_t)\cos(e^U_t), \label{eq:score-comp}
\end{align}
that is, $\mathbf u_t=\mathbf D^{-1}\nabla_t$ with $\mathbf D=\mathrm{diag}(\upsilon_L,\upsilon_U)$.
Since $\upsilon_b$ is constant, the scaling only changes the units of the gain,
$\mathbf K_{\text{scaled}}=\mathbf K_{\text{unscaled}}\mathbf D$, and leaves the model unchanged. 
We define below the IN-DCS model as follows,
\begin{equation}\label{eq:vector}
  \bm{\mu}_{t+1|t} = \mathbf{\Phi \bm{\mu}_{t|t-1} }+ \mathbf{K u_t}, \qquad {\bm{\mu}_{1|0}} =  0,
\end{equation}
where
\begin{equation*}\label{eq:matrices}
  \bm{\Phi} = \begin{pmatrix} \phi_L & 0 \\ 0 & \phi_U \end{pmatrix},
  \qquad
  \bm{K} = \begin{pmatrix} \kappa_L & 0 \\ 0 & \kappa_U \end{pmatrix}.
\end{equation*}

Both $\bm{\Phi}$ and $\bm{K}$ are diagonal matrices, and each boundary's own autoregressive
term and score continue to contribute only to that boundary's filter with no cross-lag terms. The dependence between the two boundaries is
characterised by $\bm{u}_t$ through  $\lambda$. All temporal dependence between the two boundaries is
therefore expressed via the score rather than imposed through the autoregressive
coefficient matrix. As discussed in \citep{harvey2023regime}, the score captures the meaningful interaction implied by the joint density's structure. A fully
parameterised, cross-lag specification remains a natural extension, although we note
that linear multivariate score-driven location models with a full matrix encounter
convergence difficulties in practice \citep{blazsek2021}. Hence, we do not introduce additional cross-lag parameters since the estimation may become unstable \cite{blazsek2021}. 

Setting $\lambda = 0$ in \eqref{eq:bivm} reduces
the model into two independent univariate filters, one for each
boundary, hence in this case for each boundary $b \in \{L, U\}$,
the IN-DCS approach is defined as
\begin{equation}\label{eq:obs-ind}
  \theta_{bt} = \mu_b + \mu_{bt|t-1} + e^{b}_{t},
  \qquad e^{b}_{t} \sim \mathrm{vM}(0, \upsilon_b),
\end{equation}
\begin{equation}\label{eq:filter-ind}
  \mu_{bt+1|t} = \phi_b \mu_{bt|t-1} + \kappa_b u_{bt},
  \qquad u_{bt} = \sin\!\left(\theta_{bt} - \mu_b - \mu_{bt|t-1}\right),
\end{equation}
with $\mu_{b1|0} = 0$. The parameter vector is given as
$\psi_b = (\kappa_b, \phi_b, \mu_b, \upsilon_b)'$.

 \subsection{Maximum likelihood estimation}\label{sec:mle}
In this section, we develop maximum likelihood estimation procedure. We follow the estimation approach of
\citep{harvey2024modelling}, developed for a univariate circular time series, and adapt it to interval-valued observations.

Since
the two boundary processes are coupled through the dependence parameter $\lambda$ in
the density \eqref{eq:bivm}, the score driving each boundary's own filter, given in
\eqref{eq:score-comp}, depends on both concentration parameters and on $\lambda$
jointly. All parameters of the
dependent model must be estimated jointly by maximising the full conditional
likelihood. Let the parameters be
\begin{equation*}\label{eq:psi}
 \bm{\psi} = \left(\kappa_L, \phi_L, \mu_L, \kappa_U, \phi_U, \mu_U,
    \upsilon_L, \upsilon_U, \lambda\right)'.
\end{equation*}
The
log-likelihood is given by
\begin{equation*}\label{eq:lik}
  \ell_T(\bm{\psi}) = \sum_{t=1}^{T} \Big[ -\ln C(\upsilon_L, \upsilon_U, \lambda)
    + \upsilon_L \cos\!\left(e^{L}_{t}\right)
    + \upsilon_U \cos\!\left(e^{U}_{t}\right)
    + \lambda \sin\!\left(e^{L}_{t}\right) \sin\!\left(e^{U}_{t}\right) \Big],
\end{equation*}
and the maximum likelihood estimators, for the parameter space $\bm{\Psi}$ 
\begin{equation}\label{eq:mle}
  \bm{\hat{\psi}} = \arg\max_{\bm{\psi}\in \bm\Psi} L_T(\psi).
\end{equation}
The estimation of \eqref{eq:mle} is
numerically more demanding than its independent version when $\lambda=0$. To ensure that the concentration parameters satisfy $\upsilon_L > 0$ and
$\upsilon_U > 0$, we let $\upsilon_L = \exp(\xi_L)$ and $\upsilon_U = \exp(\xi_U)$,
and use the restriction $\lambda^2 < \upsilon_L \upsilon_U$ through
\begin{equation*}
  \lambda = \sqrt{\upsilon_L \upsilon_U} \, \tanh(\gamma), \qquad \gamma \in \mathbb{R},
\end{equation*}
so that the unconstrained parameters $(\xi_L, \xi_U, \gamma)$ replace
$(\upsilon_L, \upsilon_U, \lambda)$ during numerical optimisation. This
reparameterisation allows the optimisation to be carried out over an unconstrained
space. The covariance matrix of $\hat{\bm{\psi}}$ may be estimated
from the inverse observed information \citep{harvey2023regime},
\begin{equation*}\label{eq:cov}
  \mathrm{Var}\!\left(\hat{\bm{\psi}}\right)
  = \left[ -\frac{\partial^2 L_T(\bm{\psi})}{\partial \bm{\psi} \, \partial \bm{\psi}'}
    \bigg|_{\bm{\psi} = \hat{\bm{\psi}}} \right]^{-1}.
\end{equation*}
\noindent When $\lambda=0$, for each boundary $b \in \{L, U\}$, the dynamic parameters
$\bm{\psi}_b = (\kappa_b, \phi_b, \mu_b, \upsilon_b)'$ can be estimated by maximising the log likelihood of a von Mises distribution. For invertibility and regularity conditions, see \cite{harvey2024modelling}. 

\section{Simulation Study}
\label{sec:simulation}

\begin{figure}[H]
    \centering
    \begin{subfigure}[t]{0.48\textwidth}
        \centering
        \includegraphics[width=\linewidth]{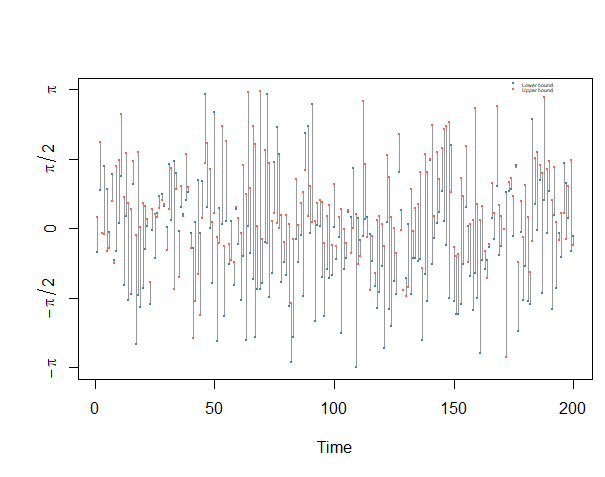}
    \end{subfigure}\hfill
    \begin{subfigure}[t]{0.48\textwidth}
        \centering
        \includegraphics[width=\linewidth]{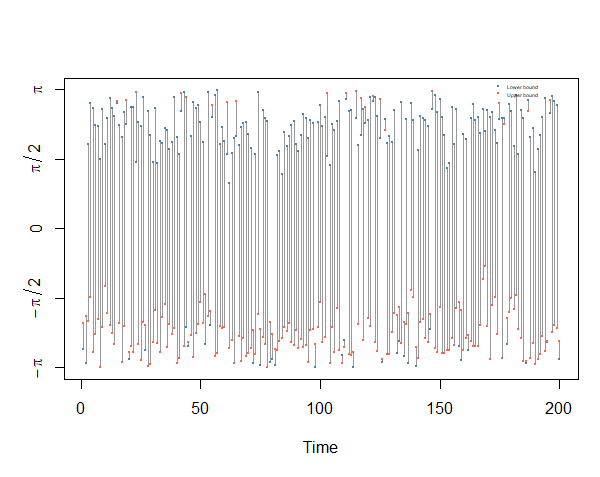}
    \end{subfigure}

    \vspace{0.5em}

    \begin{subfigure}[t]{0.47\textwidth}
        \centering
        \includegraphics[width=\linewidth]{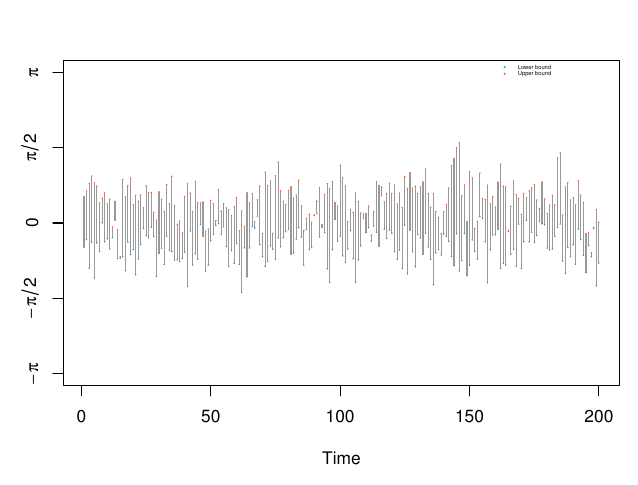}
    \end{subfigure}\hfill
    \begin{subfigure}[t]{0.48\textwidth}
        \centering
        \includegraphics[width=\linewidth]{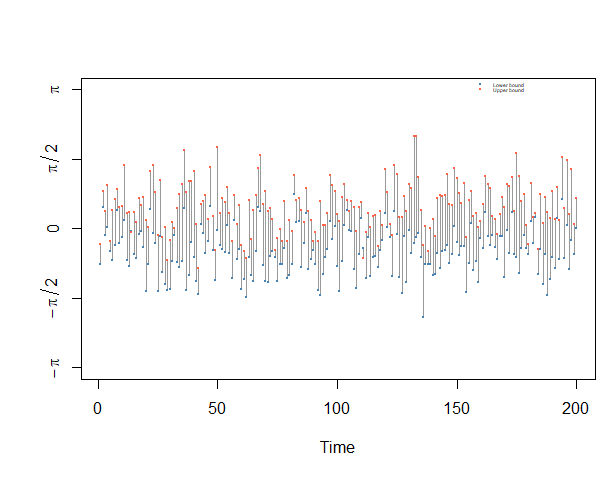}
    \end{subfigure}

 \caption{Simulated interval-valued datasets for settings H, K, C and N, representing small and wide arcs near circular cut.}
    \label{fig:study1}
\end{figure}

\begin{figure}[H]
    \centering
    \includegraphics[width=\linewidth]{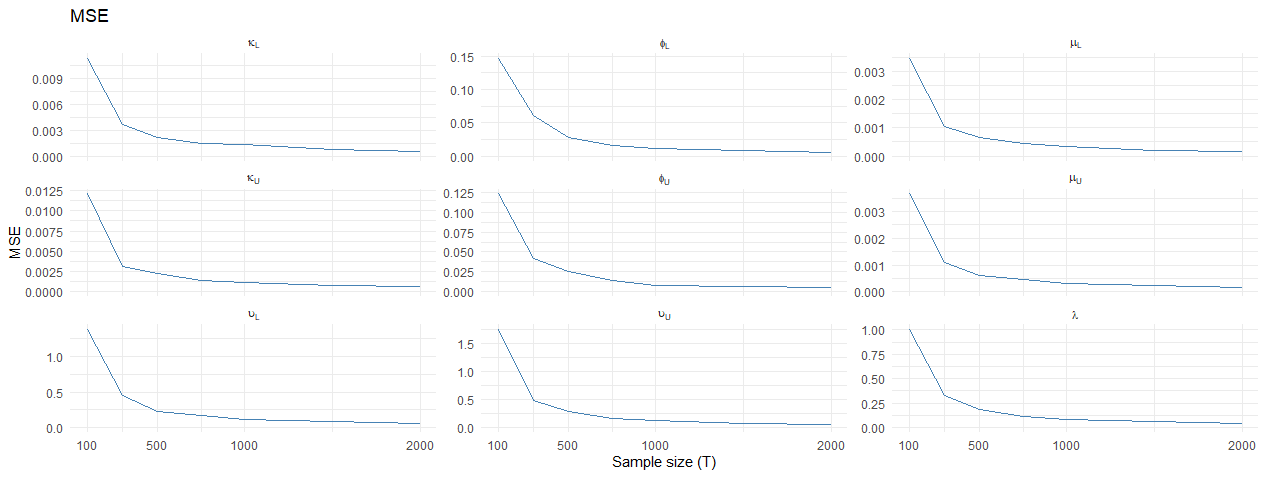}
    \caption{MSE plotted for different sample sizes using parameter setting A. }
    \label{fig:mse_study1}
\end{figure}

\begin{figure}[H]
    \centering
    \includegraphics[width=\linewidth]{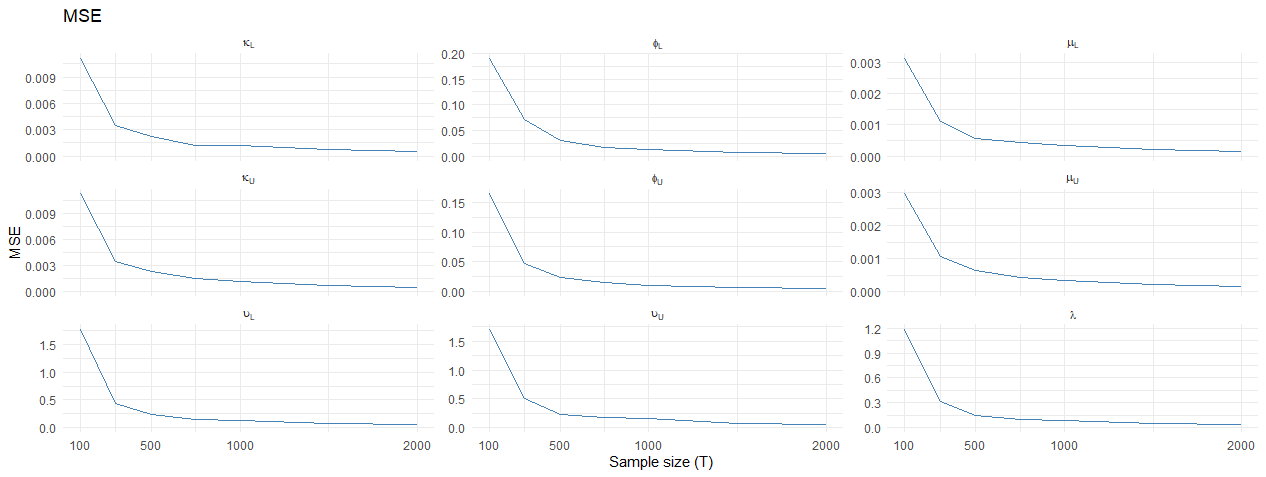}
    \caption{MSE plotted for different sample sizes using parameter setting E.}
    \label{fig:mse_study2}
\end{figure}

\begin{figure}[H]
    \centering
    \includegraphics[width=\linewidth]{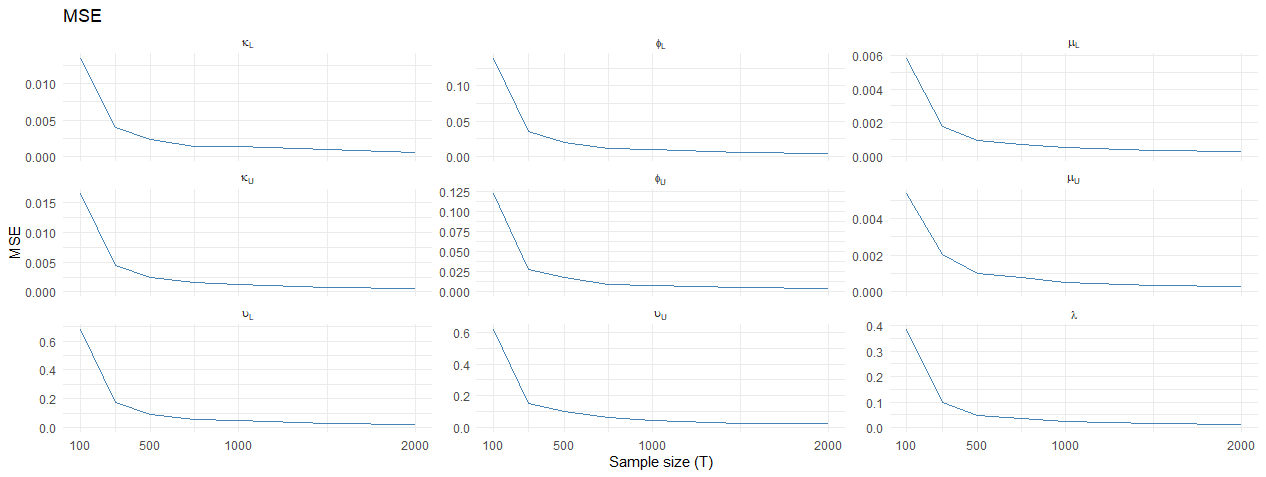}
    \caption{MSE plotted for different sample sizes using parameter setting I.}
    \label{fig:mse_study3}
\end{figure}

In this section, we conduct a simulation study to assess the performance of the maximum likelihood estimators under different parameter settings. In every setting, the data is simulated with sample sizes of $T\in\{100,500,1000\}$ and we repeated the process $1000$ times. For general scenarios were considered as below
\begin{itemize}
    \item {\bf Scenario 1}: At first we consider, four parameter settings of A, B, C, D where $\lambda=0$, the true parameter settings of A, B,C, D and the results, including  estimates, standard deviations (SD), mean squared error (MSE) and absolute bias (ABIAS) are reported in Table \ref{tab:simulation_standard}.
    \item {\bf Scenario 2}: Secondly, we consider four parameter settings of E, F, G, H where $\lambda=0.75,0.75,4,4$, respectively. The true parameter settings of E, F, G, H and the results of these simulations are reported in Table \ref{tab:simulation_lambda}.
    \item {\bf Scenario 3}: Thirdly, we consider one parameter setting of $\kappa_L=0.2, \phi_L=0.6, \mu_L=-0.5, \kappa_U=0.25, \phi_U=0.55, \mu_U=0.5,
    \upsilon_L=8, \upsilon_U=8$  where the mean directions are concentrated on a narrow arc and consider four different dependence parameters of $\lambda=0,3,-3,7$. The results of this configuration are given in Table \ref{tab:simulation_lambda_1} under settings of I, G, K and L.
    \item {\bf Scenario 4}: Lastly, we again consider one parameter setting of  $\kappa_L=0.3, \phi_L=0.5, \mu_L=2.54, \kappa_U=0.35, \phi_U=0.45, \mu_U=-2.54,
    \upsilon_L=5, \upsilon_U=5$ where the mean directions are chosen to be near the circular cut whilst the dependence parameter is again varied as $\lambda=0.3,2,-5,4.4$. The results of these simulations are reported in Table \ref{tab:simulation_lambda_2} under settings of M, N, O and P.
\end{itemize}
For illustration, the simulated upper and lower bounds, $n=200$, have been plotted in Figures \ref{fig:study1}, for setting H,K,C and N. In addition, the MSEs of the  estimators for various sample sizes are shown in \ref{fig:mse_study1}-\ref{fig:mse_study3} for parameter settings of A, E and I.
In Table \ref{tab:simulation_standard}-\ref{tab:simulation_lambda_2}, the ABIAS and MSE decrease substantially as sample size increases for all parameters, which is further visualised in the MSE plots in Figures \ref{fig:mse_study1}-\ref{fig:mse_study3}. In addition, the results in Tables \ref{tab:simulation_standard}-\ref{tab:simulation_lambda_2} indicate that the estimators perform well when the intervals are concentrated on a narrow arc, when intervals are spread over a wide arc, or located near the circular cut. The IN-DCS model addresses the limitation of the existing approach of \cite{mohan2026cinar}, which uses the link-function, which is, however, not fully rotationally invariant. In conclusion, the simulation study provides evidence of the robustness of the proposed estimation procedure. These findings support the numerical reliability of the proposed estimators over the range of independent and dependent circular interval-valued time series processes.

\begin{table}[H]
\centering
\caption{Parameter estimates, ABIAS, SD and MSE for Scenario 1.}
\label{tab:simulation_standard}
\scriptsize
\setlength{\tabcolsep}{3pt}
\renewcommand{\arraystretch}{1.1}
%
\begin{subtable}[t]{0.49\textwidth}
\centering
\begin{adjustbox}{max width=\linewidth}
\begin{tabular}{|c c S[table-format=-1.4] S[table-format=-1.4]
                    S[table-format=<1.4] S[table-format=1.4] S[table-format=1.4]|}
\hline
$T$ & Parameter & {True} & {Estimate} & {ABIAS} & {SD} & {MSE} \\
\hline
\multicolumn{7}{|c|}{A} \\
\hline
   100  & $\kappa_L$   &  0.2000 &  0.1959 & 0.0041 & 0.0753 & 0.0057 \\
     & $\kappa_U$   &  0.2500 &  0.2443 & 0.0057 & 0.0746 & 0.0056 \\
     & $\phi_L$     &  0.6000 &  0.5150 & 0.0850 & 0.2673 & 0.0786 \\
     & $\phi_U$     &  0.5500 &  0.4861 & 0.0639 & 0.2505 & 0.0668 \\
     & $\mu_L$      & -0.5000 & -0.4987 & 0.0013 & 0.0386 & 0.0015 \\
     & $\mu_U$      &  0.5000 &  0.4994 & 0.0006 & 0.0387 & 0.0015 \\
     & $\upsilon_L$ &  8.0000 &  8.1976 & 0.1976 & 0.7987 & 0.6762 \\
     & $\upsilon_U$ &  8.0000 &  8.1668 & 0.1668 & 0.7978 & 0.6636 \\
\cline{1-7}
500     & $\kappa_L$   &  0.2000 &  0.1972 & 0.0028 & 0.0456 & 0.0021 \\
     & $\kappa_U$   &  0.2500 &  0.2489 & 0.0011 & 0.0480 & 0.0023 \\
     & $\phi_L$     &  0.6000 &  0.5549 & 0.0451 & 0.1761 & 0.0330 \\
     & $\phi_U$     &  0.5500 &  0.5281 & 0.0219 & 0.1477 & 0.0223 \\
     & $\mu_L$      & -0.5000 & -0.4993 & 0.0007 & 0.0241 & 0.0006 \\
     & $\mu_U$      &  0.5000 &  0.5010 & 0.0010 & 0.0257 & 0.0007 \\
     & $\upsilon_L$ &  8.0000 &  8.0788 & 0.0788 & 0.5127 & 0.2688 \\
     & $\upsilon_U$ &  8.0000 &  8.0935 & 0.0935 & 0.4838 & 0.2425 \\
\cline{1-7}
  1000  & $\kappa_L$   &  0.2000 &  0.1994 & 0.0006 & 0.0316 & 0.0010 \\
     & $\kappa_U$   &  0.2500 &  0.2490 & 0.0010 & 0.0318 & 0.0010 \\
     & $\phi_L$     &  0.6000 &  0.5783 & 0.0217 & 0.1111 & 0.0128 \\
     & $\phi_U$     &  0.5500 &  0.5376 & 0.0124 & 0.0965 & 0.0095 \\
     & $\mu_L$      & -0.5000 & -0.5001 & 0.0001 & 0.0171 & 0.0003 \\
     & $\mu_U$      &  0.5000 &  0.4996 & 0.0004 & 0.0176 & 0.0003 \\
     & $\upsilon_L$ &  8.0000 &  8.0393 & 0.0393 & 0.3370 & 0.1149 \\
     & $\upsilon_U$ &  8.0000 &  8.0343 & 0.0343 & 0.3591 & 0.1300 \\
\hline
\multicolumn{7}{|c|}{B} \\
\hline
     100& $\kappa_L$   &  0.4000 &  0.3910 & 0.0090 & 0.0755 & 0.0058 \\
     & $\kappa_U$   &  0.5000 &  0.4897 & 0.0103 & 0.0780 & 0.0062 \\
     & $\phi_L$     &  0.7000 &  0.6591 & 0.0409 & 0.1315 & 0.0189 \\
     & $\phi_U$     &  0.6000 &  0.5727 & 0.0273 & 0.1213 & 0.0154 \\
     & $\mu_L$      & -0.5000 & -0.5024 & 0.0024 & 0.0582 & 0.0034 \\
     & $\mu_U$      &  0.5000 &  0.4977 & 0.0023 & 0.0588 & 0.0035 \\
     & $\upsilon_L$ &  8.0000 &  8.2240 & 0.2240 & 0.7949 & 0.6813 \\
     & $\upsilon_U$ &  8.0000 &  8.1692 & 0.1692 & 0.7928 & 0.6564 \\
\cline{1-7}
  500   & $\kappa_L$   &  0.4000 &  0.3972 & 0.0028 & 0.0468 & 0.0022 \\
     & $\kappa_U$   &  0.5000 &  0.4984 & 0.0016 & 0.0482 & 0.0023 \\
     & $\phi_L$     &  0.7000 &  0.6896 & 0.0104 & 0.0686 & 0.0048 \\
     & $\phi_U$     &  0.6000 &  0.5891 & 0.0109 & 0.0734 & 0.0055 \\
     & $\mu_L$      & -0.5000 & -0.4992 & 0.0008 & 0.0355 & 0.0013 \\
     & $\mu_U$      &  0.5000 &  0.5011 & 0.0011 & 0.0356 & 0.0013 \\
     & $\upsilon_L$ &  8.0000 &  8.0490 & 0.0490 & 0.4767 & 0.2293 \\
     & $\upsilon_U$ &  8.0000 &  8.0785 & 0.0785 & 0.4897 & 0.2457 \\
\cline{1-7}
  1000   & $\kappa_L$   &  0.4000 &  0.3974 & 0.0026 & 0.0330 & 0.0011 \\
     & $\kappa_U$   &  0.5000 &  0.4989 & 0.0011 & 0.0345 & 0.0012 \\
     & $\phi_L$     &  0.7000 &  0.6910 & 0.0090 & 0.0513 & 0.0027 \\
     & $\phi_U$     &  0.6000 &  0.5937 & 0.0063 & 0.0527 & 0.0028 \\
     & $\mu_L$      & -0.5000 & -0.5006 & 0.0006 & 0.0265 & 0.0007 \\
     & $\mu_U$      &  0.5000 &  0.4996 & 0.0004 & 0.0247 & 0.0006 \\
     & $\upsilon_L$ &  8.0000 &  8.0217 & 0.0217 & 0.3331 & 0.1113 \\
     & $\upsilon_U$ &  8.0000 &  8.0377 & 0.0377 & 0.3512 & 0.1246 \\
\hline
\end{tabular}
\end{adjustbox}
\end{subtable}%
\begin{subtable}[t]{0.49\textwidth}
\centering
\begin{adjustbox}{max width=\linewidth}
\begin{tabular}{|c c S[table-format=-1.4] S[table-format=-1.4]
                    S[table-format=<1.4] S[table-format=1.4] S[table-format=1.4]|}
\hline
$T$ & Parameter & {True} & {Estimate} & {ABIAS} & {SD} & {MSE} \\
\hline
\multicolumn{7}{|c|}{C} \\
\hline
  100   & $\kappa_L$   &  0.3000 &  0.2961 & 0.0039 & 0.2060 & 0.0424 \\
     & $\kappa_U$   &  0.3000 &  0.2925 & 0.0075 & 0.0708 & 0.0051 \\
     & $\phi_L$     &  0.9800 &  0.9530 & 0.0270 & 0.0597 & 0.0043 \\
     & $\phi_U$     &  0.9700 &  0.9444 & 0.0256 & 0.0579 & 0.0040 \\
     & $\mu_L$      & -0.5000 & -0.4977 & 0.0023 & 0.7732 & 0.5971 \\
     & $\mu_U$      &  0.5000 &  0.4844 & 0.0156 & 0.6321 & 0.3993 \\
     & $\upsilon_L$ &  2.0000 &  2.0319 & 0.0319 & 0.2115 & 0.0457 \\
     & $\upsilon_U$ &  2.0000 &  2.0398 & 0.0398 & 0.1746 & 0.0320 \\
\cline{1-7}
 500    & $\kappa_L$   &  0.3000 &  0.3048 & 0.0048 & 0.0961 & 0.0092 \\
     & $\kappa_U$   &  0.3000 &  0.2988 & 0.0012 & 0.0469 & 0.0022 \\
     & $\phi_L$     &  0.9800 &  0.9725 & 0.0075 & 0.0254 & 0.0007 \\
     & $\phi_U$     &  0.9700 &  0.9628 & 0.0072 & 0.0211 & 0.0005 \\
     & $\mu_L$      & -0.5000 & -0.4929 & 0.0071 & 0.6842 & 0.4676 \\
     & $\mu_U$      &  0.5000 &  0.4695 & 0.0305 & 0.4979 & 0.2485 \\
     & $\upsilon_L$ &  2.0000 &  2.0056 & 0.0056 & 0.1295 & 0.0168 \\
     & $\upsilon_U$ &  2.0000 &  2.0093 & 0.0093 & 0.1151 & 0.0133 \\
\cline{1-7}
  1000   & $\kappa_L$   &  0.3000 &  0.2981 & 0.0019 & 0.0393 & 0.0015 \\
     & $\kappa_U$   &  0.3000 &  0.2997 & 0.0003 & 0.0398 & 0.0016 \\
     & $\phi_L$     &  0.9800 &  0.9775 & 0.0025 & 0.0107 & 0.0001 \\
     & $\phi_U$     &  0.9700 &  0.9667 & 0.0033 & 0.0120 & 0.0002 \\
     & $\mu_L$      & -0.5000 & -0.5110 & 0.0110 & 0.5404 & 0.2918 \\
     & $\mu_U$      &  0.5000 &  0.4922 & 0.0078 & 0.3511 & 0.1231 \\
     & $\upsilon_L$ &  2.0000 &  2.0057 & 0.0057 & 0.0800 & 0.0064 \\
     & $\upsilon_U$ &  2.0000 &  2.0058 & 0.0058 & 0.0798 & 0.0064 \\
\hline
\multicolumn{7}{|c|}{D} \\
\hline
100     & $\kappa_L$   &  0.2000 &  0.2044 & 0.0044 & 0.1597 & 0.0255 \\
     & $\kappa_U$   &  0.2500 &  0.2422 & 0.0078 & 0.1699 & 0.0289 \\
     & $\phi_L$     &  0.5000 &  0.3431 & 0.1569 & 0.4515 & 0.2282 \\
     & $\phi_U$     &  0.4500 &  0.3363 & 0.1137 & 0.4294 & 0.1971 \\
     & $\mu_L$      & -0.5000 & -0.4964 & 0.0036 & 0.1296 & 0.0168 \\
     & $\mu_U$      &  0.5000 &  0.4990 & 0.0010 & 0.1296 & 0.0168 \\
     & $\upsilon_L$ &  1.0000 &  1.0319 & 0.0319 & 0.1195 & 0.0153 \\
     & $\upsilon_U$ &  1.0000 &  1.0220 & 0.0220 & 0.1161 & 0.0139 \\
\cline{1-7}
500     & $\kappa_L$   &  0.2000 &  0.2032 & 0.0032 & 0.0922 & 0.0085 \\
     & $\kappa_U$   &  0.2500 &  0.2515 & 0.0015 & 0.1011 & 0.0102 \\
     & $\phi_L$     &  0.5000 &  0.4259 & 0.0741 & 0.3588 & 0.1341 \\
     & $\phi_U$     &  0.4500 &  0.3776 & 0.0724 & 0.3451 & 0.1242 \\
     & $\mu_L$      & -0.5000 & -0.5034 & 0.0034 & 0.0796 & 0.0063 \\
     & $\mu_U$      &  0.5000 &  0.5061 & 0.0061 & 0.0837 & 0.0070 \\
     & $\upsilon_L$ &  1.0000 &  1.0056 & 0.0056 & 0.0761 & 0.0058 \\
     & $\upsilon_U$ &  1.0000 &  1.0113 & 0.0113 & 0.0783 & 0.0062 \\
\cline{1-7}
 1000    & $\kappa_L$   &  0.2000 &  0.2024 & 0.0024 & 0.0678 & 0.0046 \\
     & $\kappa_U$   &  0.2500 &  0.2499 & 0.0001 & 0.0719 & 0.0052 \\
     & $\phi_L$     &  0.5000 &  0.4447 & 0.0553 & 0.2772 & 0.0798 \\
     & $\phi_U$     &  0.4500 &  0.4052 & 0.0448 & 0.2425 & 0.0607 \\
     & $\mu_L$      & -0.5000 & -0.4985 & 0.0015 & 0.0547 & 0.0030 \\
     & $\mu_U$      &  0.5000 &  0.5018 & 0.0018 & 0.0567 & 0.0032 \\
     & $\upsilon_L$ &  1.0000 &  1.0024 & 0.0024 & 0.0514 & 0.0026 \\
     & $\upsilon_U$ &  1.0000 &  1.0063 & 0.0063 & 0.0508 & 0.0026 \\
\hline
\end{tabular}
\end{adjustbox}
\end{subtable} 
\label{tab:simulation_standard}
\end{table}
\begin{table}[H]
\centering
\caption{Parameter estimates, ABIAS, SD and MSE for Scenario 2.}
\label{tab:simulation_lambda}
\scriptsize
\setlength{\tabcolsep}{3pt}
\renewcommand{\arraystretch}{1.1}
%
\begin{subtable}[t]{0.49\textwidth}
\centering
\begin{adjustbox}{max width=\linewidth}
\begin{tabular}{|c c S[table-format=-1.4] S[table-format=-1.4]
                    S[table-format=<1.4] S[table-format=1.4] S[table-format=1.4]|}
\hline
$T$ & Parameter & {True} & {Estimate} & {ABIAS} & {SD} & {MSE} \\
\hline
\multicolumn{7}{|c|}{E} \\
\hline
100  & $\kappa_L$   &  0.4500 &  0.4380 & 0.0120 & 0.2316 & 0.0537 \\
     & $\phi_L$     &  0.5000 &  0.3930 & 0.1070 & 0.3147 & 0.1103 \\
     & $\mu_L$      & -0.5000 & -0.5010 & 0.0010 & 0.1843 & 0.0339 \\
     & $\kappa_U$   &  0.4800 &  0.4740 & 0.0060 & 0.2362 & 0.0557 \\
     & $\phi_U$     &  0.5000 &  0.4199 & 0.0801 & 0.2969 & 0.0945 \\
     & $\mu_U$      &  0.5000 &  0.5131 & 0.0131 & 0.1909 & 0.0366 \\
     & $\upsilon_L$ &  1.5000 &  1.5614 & 0.0614 & 0.2241 & 0.0539 \\
     & $\upsilon_U$ &  1.5000 &  1.5480 & 0.0480 & 0.2150 & 0.0485 \\
     & $\lambda$    &  0.7500 &  0.7801 & 0.0301 & 0.2595 & 0.0681 \\
\cline{1-7}
500  & $\kappa_L$   &  0.4500 &  0.4485 & 0.0015 & 0.0677 & 0.0046 \\
     & $\phi_L$     &  0.5000 &  0.4792 & 0.0208 & 0.1266 & 0.0164 \\
     & $\mu_L$      & -0.5000 & -0.4976 & 0.0024 & 0.0703 & 0.0049 \\
     & $\kappa_U$   &  0.4800 &  0.4753 & 0.0047 & 0.0718 & 0.0052 \\
     & $\phi_U$     &  0.5000 &  0.4823 & 0.0177 & 0.1129 & 0.0130 \\
     & $\mu_U$      &  0.5000 &  0.5009 & 0.0009 & 0.0741 & 0.0055 \\
     & $\upsilon_L$ &  1.5000 &  1.5185 & 0.0185 & 0.0906 & 0.0085 \\
     & $\upsilon_U$ &  1.5000 &  1.5101 & 0.0101 & 0.0908 & 0.0083 \\
     & $\lambda$    &  0.7500 &  0.7612 & 0.0112 & 0.1070 & 0.0116 \\
\cline{1-7}
1000 & $\kappa_L$   &  0.4500 &  0.4477 & 0.0023 & 0.0471 & 0.0022 \\
     & $\phi_L$     &  0.5000 &  0.4890 & 0.0110 & 0.0854 & 0.0074 \\
     & $\mu_L$      & -0.5000 & -0.4986 & 0.0014 & 0.0480 & 0.0023 \\
     & $\kappa_U$   &  0.4800 &  0.4791 & 0.0009 & 0.0493 & 0.0024 \\
     & $\phi_U$     &  0.5000 &  0.4927 & 0.0073 & 0.0774 & 0.0060 \\
     & $\mu_U$      &  0.5000 &  0.5011 & 0.0011 & 0.0499 & 0.0025 \\
     & $\upsilon_L$ &  1.5000 &  1.5072 & 0.0072 & 0.0631 & 0.0040 \\
     & $\upsilon_U$ &  1.5000 &  1.5067 & 0.0067 & 0.0644 & 0.0042 \\
     & $\lambda$    &  0.7500 &  0.7539 & 0.0039 & 0.0741 & 0.0055 \\
\hline
\multicolumn{7}{|c|}{ F} \\
\hline
100  & $\kappa_L$   &  0.5200 &  0.5154 & 0.0046 & 0.2297 & 0.0527 \\
     & $\phi_L$     &  0.5000 &  0.3974 & 0.1026 & 0.2843 & 0.0912 \\
     & $\mu_L$      & -0.5000 & -0.5048 & 0.0048 & 0.1966 & 0.0386 \\
     & $\kappa_U$   &  0.5500 &  0.5468 & 0.0032 & 0.1679 & 0.0282 \\
     & $\phi_U$     &  0.5000 &  0.4231 & 0.0769 & 0.2793 & 0.0838 \\
     & $\mu_U$      &  0.5000 &  0.4962 & 0.0038 & 0.1792 & 0.0321 \\
     & $\upsilon_L$ &  1.5000 &  1.5624 & 0.0624 & 0.2194 & 0.0520 \\
     & $\upsilon_U$ &  1.5000 &  1.5683 & 0.0683 & 0.2061 & 0.0471 \\
     & $\lambda$    &  0.7500 &  0.7832 & 0.0332 & 0.2652 & 0.0713 \\
\cline{1-7}
500  & $\kappa_L$   &  0.5200 &  0.5173 & 0.0027 & 0.0674 & 0.0045 \\
     & $\phi_L$     &  0.5000 &  0.4836 & 0.0164 & 0.1072 & 0.0117 \\
     & $\mu_L$      & -0.5000 & -0.4994 & 0.0006 & 0.0767 & 0.0059 \\
     & $\kappa_U$   &  0.5500 &  0.5472 & 0.0028 & 0.0692 & 0.0048 \\
     & $\phi_U$     &  0.5000 &  0.4800 & 0.0200 & 0.1026 & 0.0109 \\
     & $\mu_U$      &  0.5000 &  0.4934 & 0.0066 & 0.0760 & 0.0058 \\
     & $\upsilon_L$ &  1.5000 &  1.5098 & 0.0098 & 0.0883 & 0.0079 \\
     & $\upsilon_U$ &  1.5000 &  1.5121 & 0.0121 & 0.0919 & 0.0086 \\
     & $\lambda$    &  0.7500 &  0.7540 & 0.0040 & 0.1038 & 0.0108 \\
\cline{1-7}
1000 & $\kappa_L$   &  0.5200 &  0.5189 & 0.0011 & 0.0469 & 0.0022 \\
     & $\phi_L$     &  0.5000 &  0.4970 & 0.0030 & 0.0740 & 0.0055 \\
     & $\mu_L$      & -0.5000 & -0.5005 & 0.0005 & 0.0509 & 0.0026 \\
     & $\kappa_U$   &  0.5500 &  0.5454 & 0.0046 & 0.0480 & 0.0023 \\
     & $\phi_U$     &  0.5000 &  0.4905 & 0.0095 & 0.0739 & 0.0055 \\
     & $\mu_U$      &  0.5000 &  0.4984 & 0.0016 & 0.0527 & 0.0028 \\
     & $\upsilon_L$ &  1.5000 &  1.5112 & 0.0112 & 0.0684 & 0.0048 \\
     & $\upsilon_U$ &  1.5000 &  1.5002 & 0.0002 & 0.0617 & 0.0038 \\
     & $\lambda$    &  0.7500 &  0.7526 & 0.0026 & 0.0705 & 0.0050 \\
\hline
\end{tabular}
\end{adjustbox}
\end{subtable}%
\begin{subtable}[t]{0.49\textwidth}
\centering
\begin{adjustbox}{max width=\linewidth}
\begin{tabular}{|c c S[table-format=-1.4] S[table-format=-1.4]
                    S[table-format=<1.4] S[table-format=1.4] S[table-format=1.4]|}
\hline
$T$ & Parameter & {True} & {Estimate} & {ABIAS} & {SD} & {MSE} \\
\hline
\multicolumn{7}{|c|}{G} \\
\hline
100  & $\kappa_L$   &  0.2000 &  0.1954 & 0.0046 & 0.1145 & 0.0131 \\
     & $\phi_L$     &  0.6000 &  0.4701 & 0.1299 & 0.3776 & 0.1593 \\
     & $\mu_L$      & -0.5000 & -0.5048 & 0.0048 & 0.0586 & 0.0035 \\
     & $\kappa_U$   &  0.2500 &  0.2440 & 0.0060 & 0.1134 & 0.0129 \\
     & $\phi_U$     &  0.5500 &  0.4529 & 0.0971 & 0.3132 & 0.1074 \\
     & $\mu_U$      &  0.5000 &  0.4991 & 0.0009 & 0.0617 & 0.0038 \\
     & $\upsilon_L$ &  8.0000 &  8.4372 & 0.4372 & 1.1472 & 1.5055 \\
     & $\upsilon_U$ &  8.0000 &  8.5004 & 0.5004 & 1.2748 & 1.8734 \\
     & $\lambda$    &  4.0000 &  4.2508 & 0.2508 & 1.0471 & 1.1578 \\
\cline{1-7}
500  & $\kappa_L$   &  0.2000 &  0.1993 & 0.0007 & 0.0465 & 0.0022 \\
     & $\phi_L$     &  0.6000 &  0.5697 & 0.0303 & 0.1534 & 0.0244 \\
     & $\mu_L$      & -0.5000 & -0.4999 & 0.0001 & 0.0263 & 0.0007 \\
     & $\kappa_U$   &  0.2500 &  0.2444 & 0.0056 & 0.0462 & 0.0022 \\
     & $\phi_U$     &  0.5500 &  0.5244 & 0.0256 & 0.1462 & 0.0220 \\
     & $\mu_U$      &  0.5000 &  0.4998 & 0.0002 & 0.0267 & 0.0007 \\
     & $\upsilon_L$ &  8.0000 &  8.0758 & 0.0758 & 0.4865 & 0.2422 \\
     & $\upsilon_U$ &  8.0000 &  8.0916 & 0.0916 & 0.4952 & 0.2533 \\
     & $\lambda$    &  4.0000 &  4.0669 & 0.0669 & 0.4262 & 0.1859 \\
\cline{1-7}
1000 & $\kappa_L$   &  0.2000 &  0.1975 & 0.0025 & 0.0339 & 0.0012 \\
     & $\phi_L$     &  0.6000 &  0.5869 & 0.0131 & 0.1098 & 0.0122 \\
     & $\mu_L$      & -0.5000 & -0.4993 & 0.0007 & 0.0177 & 0.0003 \\
     & $\kappa_U$   &  0.2500 &  0.2494 & 0.0006 & 0.0340 & 0.0012 \\
     & $\phi_U$     &  0.5500 &  0.5398 & 0.0102 & 0.1004 & 0.0102 \\
     & $\mu_U$      &  0.5000 &  0.5000 & <0.0001 & 0.0184 & 0.0003 \\
     & $\upsilon_L$ &  8.0000 &  8.0337 & 0.0337 & 0.3464 & 0.1210 \\
     & $\upsilon_U$ &  8.0000 &  8.0421 & 0.0421 & 0.3535 & 0.1266 \\
     & $\lambda$    &  4.0000 &  4.0249 & 0.0249 & 0.2975 & 0.0890 \\
\hline
\multicolumn{7}{|c|}{H} \\
\hline
100  & $\kappa_L$   &  0.4000 &  0.3960 & 0.0040 & 0.1118 & 0.0125 \\
     & $\phi_L$     &  0.7000 &  0.6104 & 0.0896 & 0.2077 & 0.0511 \\
     & $\mu_L$      & -0.5000 & -0.5011 & 0.0011 & 0.0928 & 0.0086 \\
     & $\kappa_U$   &  0.5000 &  0.4802 & 0.0198 & 0.1111 & 0.0127 \\
     & $\phi_U$     &  0.6000 &  0.5428 & 0.0572 & 0.2036 & 0.0447 \\
     & $\mu_U$      &  0.5000 &  0.4988 & 0.0012 & 0.0828 & 0.0068 \\
     & $\upsilon_L$ &  8.0000 &  8.4612 & 0.4612 & 1.1876 & 1.6214 \\
     & $\upsilon_U$ &  8.0000 &  8.4058 & 0.4058 & 1.1732 & 1.5394 \\
     & $\lambda$    &  4.0000 &  4.2580 & 0.2580 & 0.9632 & 0.9932 \\
\cline{1-7}
500  & $\kappa_L$   &  0.4000 &  0.3990 & 0.0010 & 0.0475 & 0.0023 \\
     & $\phi_L$     &  0.7000 &  0.6850 & 0.0150 & 0.0719 & 0.0054 \\
     & $\mu_L$      & -0.5000 & -0.5001 & 0.0001 & 0.0373 & 0.0014 \\
     & $\kappa_U$   &  0.5000 &  0.4959 & 0.0041 & 0.0480 & 0.0023 \\
     & $\phi_U$     &  0.6000 &  0.5884 & 0.0116 & 0.0734 & 0.0055 \\
     & $\mu_U$      &  0.5000 &  0.4980 & 0.0020 & 0.0361 & 0.0013 \\
     & $\upsilon_L$ &  8.0000 &  8.0573 & 0.0573 & 0.4892 & 0.2423 \\
     & $\upsilon_U$ &  8.0000 &  8.0671 & 0.0671 & 0.4771 & 0.2318 \\
     & $\lambda$    &  4.0000 &  4.0424 & 0.0424 & 0.3961 & 0.1585 \\
\cline{1-7}
1000 & $\kappa_L$   &  0.4000 &  0.3992 & 0.0008 & 0.0337 & 0.0011 \\
     & $\phi_L$     &  0.7000 &  0.6953 & 0.0047 & 0.0473 & 0.0023 \\
     & $\mu_L$      & -0.5000 & -0.4988 & 0.0012 & 0.0268 & 0.0007 \\
     & $\kappa_U$   &  0.5000 &  0.5007 & 0.0007 & 0.0330 & 0.0011 \\
     & $\phi_U$     &  0.6000 &  0.5947 & 0.0053 & 0.0540 & 0.0029 \\
     & $\mu_U$      &  0.5000 &  0.4988 & 0.0012 & 0.0256 & 0.0007 \\
     & $\upsilon_L$ &  8.0000 &  8.0639 & 0.0639 & 0.3353 & 0.1164 \\
     & $\upsilon_U$ &  8.0000 &  8.0510 & 0.0510 & 0.3494 & 0.1245 \\
     & $\lambda$    &  4.0000 &  4.0357 & 0.0357 & 0.2892 & 0.0848 \\
\hline
\end{tabular}
\end{adjustbox}
\end{subtable}
\label{tab:simulation_lambda}
\end{table}
\begin{table}[H]
\centering
\caption{Parameter estimates, ABIAS, SD and MSE for Scenario 3.}
\label{tab:simulation_lambda_1}
\scriptsize
\setlength{\tabcolsep}{3pt}
\renewcommand{\arraystretch}{1.1}
\begin{subtable}[t]{0.49\textwidth}
\centering
\begin{adjustbox}{max width=\linewidth}
\begin{tabular}{|c c S[table-format=-1.4] S[table-format=-1.4]
                    S[table-format=<1.4] S[table-format=1.4] S[table-format=1.4]|}
\hline
$T$ & Parameter & {True} & {Estimate} & {ABIAS} & {SD} & {MSE} \\
\hline
\multicolumn{7}{|c|}{I} \\
\hline
100     & $\kappa_L$   &  0.2000 &  0.2034 & 0.0034 & 0.1139 & 0.0130 \\
     & $\phi_L$     &  0.6000 &  0.4526 & 0.1474 & 0.3757 & 0.1627 \\
     & $\mu_L$      & -0.5000 & -0.5007 & 0.0007 & 0.0565 & 0.0032 \\
     & $\kappa_U$   &  0.2500 &  0.2383 & 0.0117 & 0.1141 & 0.0131 \\
     & $\phi_U$     &  0.5500 &  0.4078 & 0.1422 & 0.3757 & 0.1612 \\
     & $\mu_U$      &  0.5000 &  0.5001 & 0.0001 & 0.0528 & 0.0028 \\
     & $\upsilon_L$ &  8.0000 &  8.4064 & 0.4064 & 1.1729 & 1.5390 \\
     & $\upsilon_U$ &  8.0000 &  8.4576 & 0.4576 & 1.2623 & 1.8009 \\
     & $\lambda$    &  0.0000 & -0.0055 & 0.0055 & 0.9392 & 0.8810 \\
\cline{1-7}
 500    & $\kappa_L$   &  0.2000 &  0.1991 & 0.0009 & 0.0474 & 0.0022 \\
     & $\phi_L$     &  0.6000 &  0.5607 & 0.0393 & 0.1784 & 0.0333 \\
     & $\mu_L$      & -0.5000 & -0.5005 & 0.0005 & 0.0247 & 0.0006 \\
     & $\kappa_U$   &  0.2500 &  0.2499 & 0.0001 & 0.0482 & 0.0023 \\
     & $\phi_U$     &  0.5500 &  0.5201 & 0.0299 & 0.1443 & 0.0217 \\
     & $\mu_U$      &  0.5000 &  0.5002 & 0.0002 & 0.0247 & 0.0006 \\
     & $\upsilon_L$ &  8.0000 &  8.1089 & 0.1089 & 0.4976 & 0.2591 \\
     & $\upsilon_U$ &  8.0000 &  8.0904 & 0.0904 & 0.4782 & 0.2365 \\
     & $\lambda$    &  0.0000 & -0.0119 & 0.0119 & 0.3550 & 0.1260 \\
\cline{1-7}
 1000    & $\kappa_L$   &  0.2000 &  0.1975 & 0.0025 & 0.0334 & 0.0011 \\
     & $\phi_L$     &  0.6000 &  0.5847 & 0.0153 & 0.1149 & 0.0134 \\
     & $\mu_L$      & -0.5000 & -0.5007 & 0.0007 & 0.0163 & 0.0003 \\
     & $\kappa_U$   &  0.2500 &  0.2492 & 0.0008 & 0.0334 & 0.0011 \\
     & $\phi_U$     &  0.5500 &  0.5326 & 0.0174 & 0.1043 & 0.0112 \\
     & $\mu_U$      &  0.5000 &  0.5010 & 0.0010 & 0.0175 & 0.0003 \\
     & $\upsilon_L$ &  8.0000 &  8.0347 & 0.0347 & 0.3405 & 0.1170 \\
     & $\upsilon_U$ &  8.0000 &  8.0370 & 0.0370 & 0.3250 & 0.1068 \\
     & $\lambda$    &  0.0000 & -0.0049 & 0.0049 & 0.2564 & 0.0657 \\
\hline
\multicolumn{7}{|c|}{J} \\
\hline
 100    & $\kappa_L$   &  0.2000 &  0.1905 & 0.0095 & 0.1151 & 0.0133 \\
     & $\phi_L$     &  0.6000 &  0.4745 & 0.1255 & 0.3601 & 0.1453 \\
     & $\mu_L$      & -0.5000 & -0.4980 & 0.0020 & 0.0577 & 0.0033 \\
     & $\kappa_U$   &  0.2500 &  0.2418 & 0.0082 & 0.1112 & 0.0124 \\
     & $\phi_U$     &  0.5500 &  0.4449 & 0.1051 & 0.3557 & 0.1374 \\
     & $\mu_U$      &  0.5000 &  0.5011 & 0.0011 & 0.0550 & 0.0030 \\
     & $\upsilon_L$ &  8.0000 &  8.4717 & 0.4717 & 1.1963 & 1.6518 \\
     & $\upsilon_U$ &  8.0000 &  8.4833 & 0.4833 & 1.2140 & 1.7056 \\
     & $\lambda$    &  3.0000 &  3.2564 & 0.2564 & 0.9937 & 1.0520 \\
\cline{1-7}
500     & $\kappa_L$   &  0.2000 &  0.2012 & 0.0012 & 0.0483 & 0.0023 \\
     & $\phi_L$     &  0.6000 &  0.5592 & 0.0408 & 0.1707 & 0.0308 \\
     & $\mu_L$      & -0.5000 & -0.4998 & 0.0002 & 0.0242 & 0.0006 \\
     & $\kappa_U$   &  0.2500 &  0.2484 & 0.0016 & 0.0469 & 0.0022 \\
     & $\phi_U$     &  0.5500 &  0.5232 & 0.0268 & 0.1470 & 0.0223 \\
     & $\mu_U$      &  0.5000 &  0.5000 & <0.0001 & 0.0257 & 0.0007 \\
     & $\upsilon_L$ &  8.0000 &  8.1164 & 0.1164 & 0.4815 & 0.2451 \\
     & $\upsilon_U$ &  8.0000 &  8.1107 & 0.1107 & 0.5007 & 0.2626 \\
     & $\lambda$    &  3.0000 &  3.0715 & 0.0715 & 0.4083 & 0.1716 \\
\cline{1-7}
1000     & $\kappa_L$   &  0.2000 &  0.1977 & 0.0023 & 0.0328 & 0.0011 \\
     & $\phi_L$     &  0.6000 &  0.5825 & 0.0175 & 0.1123 & 0.0129 \\
     & $\mu_L$      & -0.5000 & -0.4994 & 0.0006 & 0.0176 & 0.0003 \\
     & $\kappa_U$   &  0.2500 &  0.2511 & 0.0011 & 0.0324 & 0.0011 \\
     & $\phi_U$     &  0.5500 &  0.5360 & 0.0140 & 0.0979 & 0.0098 \\
     & $\mu_U$      &  0.5000 &  0.4997 & 0.0003 & 0.0181 & 0.0003 \\
     & $\upsilon_L$ &  8.0000 &  8.0606 & 0.0606 & 0.3574 & 0.1313 \\
     & $\upsilon_U$ &  8.0000 &  8.0483 & 0.0483 & 0.3308 & 0.1116 \\
     & $\lambda$    &  3.0000 &  3.0308 & 0.0308 & 0.2860 & 0.0827 \\
\hline
\end{tabular}
\end{adjustbox}
\end{subtable}%
\begin{subtable}[t]{0.49\textwidth}
\centering
\begin{adjustbox}{max width=\linewidth}
\begin{tabular}{|c c S[table-format=-1.4] S[table-format=-1.4]
                    S[table-format=<1.4] S[table-format=1.4] S[table-format=1.4]|}
\hline
$T$ & Parameter & {True} & {Estimate} & {ABIAS} & {SD} & {MSE} \\
\hline
\multicolumn{7}{|c|}{K} \\
\hline
100     & $\kappa_L$   &  0.2000 &  0.1952 & 0.0048 & 0.1163 & 0.0135 \\
     & $\phi_L$     &  0.6000 &  0.4759 & 0.1241 & 0.3583 & 0.1436 \\
     & $\mu_L$      & -0.5000 & -0.5011 & 0.0011 & 0.0540 & 0.0029 \\
     & $\kappa_U$   &  0.2500 &  0.2400 & 0.0100 & 0.1165 & 0.0137 \\
     & $\phi_U$     &  0.5500 &  0.4558 & 0.0942 & 0.3388 & 0.1235 \\
     & $\mu_U$      &  0.5000 &  0.5002 & 0.0002 & 0.0563 & 0.0032 \\
     & $\upsilon_L$ &  8.0000 &  8.4179 & 0.4179 & 1.1775 & 1.5595 \\
     & $\upsilon_U$ &  8.0000 &  8.4784 & 0.4784 & 1.2615 & 1.8183 \\
     & $\lambda$    & -3.0000 & -3.1478 & 0.1478 & 0.9976 & 1.0158 \\
\cline{1-7}
500    & $\kappa_L$   &  0.2000 &  0.2002 & 0.0002 & 0.0470 & 0.0022 \\
     & $\phi_L$     &  0.6000 &  0.5662 & 0.0338 & 0.1650 & 0.0283 \\
     & $\mu_L$      & -0.5000 & -0.5011 & 0.0011 & 0.0260 & 0.0007 \\
     & $\kappa_U$   &  0.2500 &  0.2475 & 0.0025 & 0.0467 & 0.0022 \\
     & $\phi_U$     &  0.5500 &  0.5213 & 0.0287 & 0.1500 & 0.0233 \\
     & $\mu_U$      &  0.5000 &  0.5004 & 0.0004 & 0.0240 & 0.0006 \\
     & $\upsilon_L$ &  8.0000 &  8.1307 & 0.1307 & 0.5086 & 0.2754 \\
     & $\upsilon_U$ &  8.0000 &  8.0759 & 0.0759 & 0.5036 & 0.2590 \\
     & $\lambda$    & -3.0000 & -3.0389 & 0.0389 & 0.4094 & 0.1689 \\
\cline{1-7}
1000     & $\kappa_L$   &  0.2000 &  0.1993 & 0.0007 & 0.0337 & 0.0011 \\
     & $\phi_L$     &  0.6000 &  0.5829 & 0.0171 & 0.1074 & 0.0118 \\
     & $\mu_L$      & -0.5000 & -0.5008 & 0.0008 & 0.0173 & 0.0003 \\
     & $\kappa_U$   &  0.2500 &  0.2464 & 0.0036 & 0.0333 & 0.0011 \\
     & $\phi_U$     &  0.5500 &  0.5429 & 0.0071 & 0.0969 & 0.0094 \\
     & $\mu_U$      &  0.5000 &  0.5003 & 0.0003 & 0.0181 & 0.0003 \\
     & $\upsilon_L$ &  8.0000 &  8.0240 & 0.0240 & 0.3407 & 0.1165 \\
     & $\upsilon_U$ &  8.0000 &  8.0272 & 0.0272 & 0.3535 & 0.1255 \\
     & $\lambda$    & -3.0000 & -2.9973 & 0.0027 & 0.2825 & 0.0797 \\
\hline
\multicolumn{7}{|c|}{L} \\
\hline
    100 & $\kappa_L$   &  0.2000 &  0.1931 & 0.0069 & 0.1164 & 0.0136 \\
     & $\phi_L$     &  0.6000 &  0.4666 & 0.1334 & 0.3608 & 0.1478 \\
     & $\mu_L$      & -0.5000 & -0.5016 & 0.0016 & 0.0654 & 0.0043 \\
     & $\kappa_U$   &  0.2500 &  0.2384 & 0.0116 & 0.1195 & 0.0144 \\
     & $\phi_U$     &  0.5500 &  0.4759 & 0.0741 & 0.3278 & 0.1128 \\
     & $\mu_U$      &  0.5000 &  0.4995 & 0.0005 & 0.0658 & 0.0043 \\
     & $\upsilon_L$ &  8.0000 &  8.5399 & 0.5399 & 1.2076 & 1.7478 \\
     & $\upsilon_U$ &  8.0000 &  8.5017 & 0.5017 & 1.1553 & 1.5848 \\
     & $\lambda$    &  7.0000 &  7.4843 & 0.4843 & 1.2207 & 1.7229 \\
\cline{1-7}
500     & $\kappa_L$   &  0.2000 &  0.1993 & 0.0007 & 0.0468 & 0.0022 \\
     & $\phi_L$     &  0.6000 &  0.5766 & 0.0234 & 0.1753 & 0.0312 \\
     & $\mu_L$      & -0.5000 & -0.5012 & 0.0012 & 0.0268 & 0.0007 \\
     & $\kappa_U$   &  0.2500 &  0.2508 & 0.0008 & 0.0516 & 0.0027 \\
     & $\phi_U$     &  0.5500 &  0.5245 & 0.0255 & 0.1504 & 0.0233 \\
     & $\mu_U$      &  0.5000 &  0.4995 & 0.0005 & 0.0291 & 0.0008 \\
     & $\upsilon_L$ &  8.0000 &  8.1406 & 0.1406 & 0.5019 & 0.2713 \\
     & $\upsilon_U$ &  8.0000 &  8.1117 & 0.1117 & 0.4750 & 0.2378 \\
     & $\lambda$    &  7.0000 &  7.1237 & 0.1237 & 0.5252 & 0.2908 \\
\cline{1-7}
1000     & $\kappa_L$   &  0.2000 &  0.1984 & 0.0016 & 0.0327 & 0.0011 \\
     & $\phi_L$     &  0.6000 &  0.5811 & 0.0189 & 0.1146 & 0.0135 \\
     & $\mu_L$      & -0.5000 & -0.5003 & 0.0003 & 0.0198 & 0.0004 \\
     & $\kappa_U$   &  0.2500 &  0.2487 & 0.0013 & 0.0335 & 0.0011 \\
     & $\phi_U$     &  0.5500 &  0.5380 & 0.0120 & 0.0945 & 0.0091 \\
     & $\mu_U$      &  0.5000 &  0.5005 & 0.0005 & 0.0200 & 0.0004 \\
     & $\upsilon_L$ &  8.0000 &  8.0653 & 0.0653 & 0.3347 & 0.1161 \\
     & $\upsilon_U$ &  8.0000 &  8.0716 & 0.0716 & 0.3509 & 0.1281 \\
     & $\lambda$    &  7.0000 &  7.0670 & 0.0670 & 0.3637 & 0.1366 \\
\hline
\end{tabular}
\end{adjustbox}
\end{subtable}
\label{tab:simulation_lambda_1}
\end{table}
\begin{table}[H]
\centering
\caption{Parameter estimates, ABIAS, SD and MSE for Scenario 4.}
\label{tab:simulation_lambda_2}
\scriptsize
\setlength{\tabcolsep}{3pt}
\renewcommand{\arraystretch}{1.1}
%
\begin{subtable}[t]{0.49\textwidth}
\centering
\begin{adjustbox}{max width=\linewidth}
\begin{tabular}{|c c S[table-format=-1.4] S[table-format=-1.4]
                    S[table-format=<1.4] S[table-format=1.4] S[table-format=1.4]|}
\hline
$T$ & Parameter & {True} & {Estimate} & {ABIAS} & {SD} & {MSE} \\
\hline
\multicolumn{7}{|c|}{M} \\
\hline
100
     & $\kappa_L$   &  0.3000 &  0.2882 & 0.0118 & 0.1142 & 0.0132 \\
     & $\phi_L$     &  0.5000 &  0.3730 & 0.1270 & 0.3327 & 0.1267 \\
     & $\mu_L$      &  2.5416 &  2.5399 & 0.0017 & 0.0705 & 0.0050 \\
     & $\kappa_U$   &  0.3500 &  0.3397 & 0.0103 & 0.1169 & 0.0138 \\
     & $\phi_U$     &  0.4500 &  0.3441 & 0.1059 & 0.3189 & 0.1128 \\
     & $\mu_U$      & -2.5416 & -2.5472 & 0.0056 & 0.0753 & 0.0057 \\
     & $\upsilon_L$ &  5.0000 &  5.2641 & 0.2641 & 0.7291 & 0.6008 \\
     & $\upsilon_U$ &  5.0000 &  5.2927 & 0.2927 & 0.7594 & 0.6617 \\
     & $\lambda$    &  0.3000 &  0.3547 & 0.0547 & 0.5850 & 0.3448 \\
\cline{1-7}
500
     & $\kappa_L$   &  0.3000 &  0.2957 & 0.0043 & 0.0492 & 0.0024 \\
     & $\phi_L$     &  0.5000 &  0.4767 & 0.0233 & 0.1326 & 0.0181 \\
     & $\mu_L$      &  2.5416 &  2.5424 & 0.0009 & 0.0332 & 0.0011 \\
     & $\kappa_U$   &  0.3500 &  0.3447 & 0.0053 & 0.0505 & 0.0026 \\
     & $\phi_U$     &  0.4500 &  0.4292 & 0.0208 & 0.1318 & 0.0178 \\
     & $\mu_U$      & -2.5416 & -2.5412 & 0.0004 & 0.0325 & 0.0011 \\
     & $\upsilon_L$ &  5.0000 &  5.0353 & 0.0353 & 0.2961 & 0.0888 \\
     & $\upsilon_U$ &  5.0000 &  5.0527 & 0.0527 & 0.2981 & 0.0915 \\
     & $\lambda$    &  0.3000 &  0.3126 & 0.0126 & 0.2312 & 0.0535 \\
\cline{1-7}
1000
     & $\kappa_L$   &  0.3000 &  0.2968 & 0.0032 & 0.0354 & 0.0013 \\
     & $\phi_L$     &  0.5000 &  0.4916 & 0.0084 & 0.0892 & 0.0080 \\
     & $\mu_L$      &  2.5416 &  2.5422 & 0.0006 & 0.0224 & 0.0005 \\
     & $\kappa_U$   &  0.3500 &  0.3490 & 0.0010 & 0.0355 & 0.0013 \\
     & $\phi_U$     &  0.4500 &  0.4378 & 0.0122 & 0.0879 & 0.0079 \\
     & $\mu_U$      & -2.5416 & -2.5416 & <0.0001 & 0.0233 & 0.0005 \\
     & $\upsilon_L$ &  5.0000 &  5.0269 & 0.0269 & 0.2042 & 0.0424 \\
     & $\upsilon_U$ &  5.0000 &  5.0265 & 0.0265 & 0.2061 & 0.0431 \\
     & $\lambda$    &  0.3000 &  0.3178 & 0.0178 & 0.1698 & 0.0291 \\
\hline
\multicolumn{7}{|c|}{N} \\
\hline
100
     & $\kappa_L$   &  0.3000 &  0.2880 & 0.0120 & 0.1123 & 0.0127 \\
     & $\phi_L$     &  0.5000 &  0.3893 & 0.1107 & 0.3431 & 0.1299 \\
     & $\mu_L$      &  2.5416 &  2.5416 & <0.0001 & 0.0789 & 0.0062 \\
     & $\kappa_U$   &  0.3500 &  0.3354 & 0.0146 & 0.1162 & 0.0137 \\
     & $\phi_U$     &  0.4500 &  0.3554 & 0.0946 & 0.3150 & 0.1081 \\
     & $\mu_U$      & -2.5416 & -2.5459 & 0.0043 & 0.0756 & 0.0057 \\
     & $\upsilon_L$ &  5.0000 &  5.3034 & 0.3034 & 0.7609 & 0.6704 \\
     & $\upsilon_U$ &  5.0000 &  5.2841 & 0.2841 & 0.7030 & 0.5743 \\
     & $\lambda$    &  2.0000 &  2.1248 & 0.1248 & 0.6407 & 0.4256 \\
\cline{1-7}
500
     & $\kappa_L$   &  0.3000 &  0.2974 & 0.0026 & 0.0482 & 0.0023 \\
     & $\phi_L$     &  0.5000 &  0.4770 & 0.0230 & 0.1428 & 0.0209 \\
     & $\mu_L$      &  2.5416 &  2.5412 & 0.0004 & 0.0326 & 0.0011 \\
     & $\kappa_U$   &  0.3500 &  0.3450 & 0.0050 & 0.0490 & 0.0024 \\
     & $\phi_U$     &  0.4500 &  0.4322 & 0.0178 & 0.1275 & 0.0166 \\
     & $\mu_U$      & -2.5416 & -2.5417 & 0.0001 & 0.0330 & 0.0011 \\
     & $\upsilon_L$ &  5.0000 &  5.0336 & 0.0336 & 0.3006 & 0.0914 \\
     & $\upsilon_U$ &  5.0000 &  5.0528 & 0.0528 & 0.2944 & 0.0894 \\
     & $\lambda$    &  2.0000 &  2.0320 & 0.0320 & 0.2546 & 0.0658 \\
\cline{1-7}
1000     & $\kappa_L$   &  0.3000 &  0.2993 & 0.0007 & 0.0341 & 0.0012 \\
     & $\phi_L$     &  0.5000 &  0.4925 & 0.0075 & 0.0896 & 0.0081 \\
     & $\mu_L$      &  2.5416 &  2.5426 & 0.0010 & 0.0237 & 0.0006 \\
     & $\kappa_U$   &  0.3500 &  0.3483 & 0.0017 & 0.0345 & 0.0012 \\
     & $\phi_U$     &  0.4500 &  0.4371 & 0.0129 & 0.0898 & 0.0082 \\
     & $\mu_U$      & -2.5416 & -2.5411 & 0.0005 & 0.0251 & 0.0006 \\
     & $\upsilon_L$ &  5.0000 &  5.0265 & 0.0265 & 0.2037 & 0.0422 \\
     & $\upsilon_U$ &  5.0000 &  5.0212 & 0.0212 & 0.2049 & 0.0424 \\
     & $\lambda$    &  2.0000 &  2.0067 & 0.0067 & 0.1714 & 0.0294 \\
\hline
\end{tabular}
\end{adjustbox}
\end{subtable}%
\begin{subtable}[t]{0.49\textwidth}
\centering
\begin{adjustbox}{max width=\linewidth}
\begin{tabular}{|c c S[table-format=-1.4] S[table-format=-1.4]
                    S[table-format=<1.4] S[table-format=1.4] S[table-format=1.4]|}
\hline
$T$ & Parameter & {True} & {Estimate} & {ABIAS} & {SD} & {MSE} \\
\hline
\multicolumn{7}{|c|}{O} \\
\hline
100     & $\kappa_L$   &  0.3000 &  0.2974 & 0.0026 & 0.1065 & 0.0113 \\
     & $\phi_L$     &  0.5000 &  0.3842 & 0.1158 & 0.3283 & 0.1211 \\
     & $\mu_L$      &  2.5416 &  2.5465 & 0.0049 & 0.0854 & 0.0073 \\
     & $\kappa_U$   &  0.3500 &  0.3418 & 0.0082 & 0.1109 & 0.0123 \\
     & $\phi_U$     &  0.4500 &  0.3762 & 0.0738 & 0.2872 & 0.0878 \\
     & $\mu_U$      & -2.5416 & -2.5497 & 0.0081 & 0.0824 & 0.0069 \\
     & $\upsilon_L$ &  5.0000 &  5.2550 & 0.2550 & 0.6640 & 0.5054 \\
     & $\upsilon_U$ &  5.0000 &  5.2150 & 0.2150 & 0.6922 & 0.5248 \\
     & $\lambda$    & -5.0000 & -5.0755 & 0.0755 & 0.6333 & 0.4062 \\
\cline{1-7}
500     & $\kappa_L$   &  0.3000 &  0.2987 & 0.0013 & 0.0476 & 0.0023 \\
     & $\phi_L$     &  0.5000 &  0.4852 & 0.0148 & 0.1273 & 0.0164 \\
     & $\mu_L$      &  2.5416 &  2.5430 & 0.0015 & 0.0361 & 0.0013 \\
     & $\kappa_U$   &  0.3500 &  0.3515 & 0.0015 & 0.0481 & 0.0023 \\
     & $\phi_U$     &  0.4500 &  0.4366 & 0.0134 & 0.1185 & 0.0142 \\
     & $\mu_U$      & -2.5416 & -2.5432 & 0.0017 & 0.0345 & 0.0012 \\
     & $\upsilon_L$ &  5.0000 &  5.0110 & 0.0110 & 0.3022 & 0.0913 \\
     & $\upsilon_U$ &  5.0000 &  5.0126 & 0.0126 & 0.2866 & 0.0822 \\
     & $\lambda$    & -5.0000 & -4.9398 & 0.0602 & 0.2824 & 0.0832 \\
\cline{1-7}
1000     & $\kappa_L$   &  0.3000 &  0.3009 & 0.0009 & 0.0334 & 0.0011 \\
     & $\phi_L$     &  0.5000 &  0.4907 & 0.0093 & 0.0883 & 0.0079 \\
     & $\mu_L$      &  2.5416 &  2.5423 & 0.0007 & 0.0245 & 0.0006 \\
     & $\kappa_U$   &  0.3500 &  0.3494 & 0.0006 & 0.0336 & 0.0011 \\
     & $\phi_U$     &  0.4500 &  0.4475 & 0.0025 & 0.0831 & 0.0069 \\
     & $\mu_U$      & -2.5416 & -2.5399 & 0.0017 & 0.0257 & 0.0007 \\
     & $\upsilon_L$ &  5.0000 &  5.0052 & 0.0052 & 0.1999 & 0.0400 \\
     & $\upsilon_U$ &  5.0000 &  5.0135 & 0.0135 & 0.2030 & 0.0413 \\
     & $\lambda$    & -5.0000 & -4.9625 & 0.0375 & 0.1880 & 0.0367 \\
\hline
\multicolumn{7}{|c|}{P} \\
\hline
100     & $\kappa_L$   &  0.3000 &  0.2886 & 0.0114 & 0.1186 & 0.0142 \\
     & $\phi_L$     &  0.5000 &  0.3927 & 0.1073 & 0.3176 & 0.1122 \\
     & $\mu_L$      &  2.5416 &  2.5411 & 0.0005 & 0.0786 & 0.0062 \\
     & $\kappa_U$   &  0.3500 &  0.3370 & 0.0130 & 0.1133 & 0.0130 \\
     & $\phi_U$     &  0.4500 &  0.3647 & 0.0853 & 0.3047 & 0.1000 \\
     & $\mu_U$      & -2.5416 & -2.5394 & 0.0022 & 0.0847 & 0.0072 \\
     & $\upsilon_L$ &  5.0000 &  5.3015 & 0.3015 & 0.7371 & 0.6335 \\
     & $\upsilon_U$ &  5.0000 &  5.3127 & 0.3127 & 0.7404 & 0.6452 \\
     & $\lambda$    &  4.4000 &  4.6698 & 0.2698 & 0.7679 & 0.6617 \\
\cline{1-7}
500     & $\kappa_L$   &  0.3000 &  0.3007 & 0.0007 & 0.0496 & 0.0025 \\
     & $\phi_L$     &  0.5000 &  0.4827 & 0.0173 & 0.1348 & 0.0184 \\
     & $\mu_L$      &  2.5416 &  2.5433 & 0.0017 & 0.0383 & 0.0015 \\
     & $\kappa_U$   &  0.3500 &  0.3438 & 0.0062 & 0.0498 & 0.0025 \\
     & $\phi_U$     &  0.4500 &  0.4457 & 0.0043 & 0.1225 & 0.0150 \\
     & $\mu_U$      & -2.5416 & -2.5415 & 0.0001 & 0.0352 & 0.0012 \\
     & $\upsilon_L$ &  5.0000 &  5.0591 & 0.0591 & 0.2968 & 0.0915 \\
     & $\upsilon_U$ &  5.0000 &  5.0768 & 0.0768 & 0.2934 & 0.0919 \\
     & $\lambda$    &  4.4000 &  4.4804 & 0.0804 & 0.3319 & 0.1165 \\
\cline{1-7}
1000     & $\kappa_L$   &  0.3000 &  0.2963 & 0.0037 & 0.0342 & 0.0012 \\
     & $\phi_L$     &  0.5000 &  0.4929 & 0.0071 & 0.0930 & 0.0087 \\
     & $\mu_L$      &  2.5416 &  2.5412 & 0.0004 & 0.0256 & 0.0007 \\
     & $\kappa_U$   &  0.3500 &  0.3495 & 0.0005 & 0.0343 & 0.0012 \\
     & $\phi_U$     &  0.4500 &  0.4525 & 0.0025 & 0.0808 & 0.0065 \\
     & $\mu_U$      & -2.5416 & -2.5423 & 0.0007 & 0.0262 & 0.0007 \\
     & $\upsilon_L$ &  5.0000 &  5.0312 & 0.0312 & 0.2090 & 0.0446 \\
     & $\upsilon_U$ &  5.0000 &  5.0408 & 0.0408 & 0.2121 & 0.0466 \\
     & $\lambda$    &  4.4000 &  4.4340 & 0.0340 & 0.2326 & 0.0552 \\
\hline
\end{tabular}
\end{adjustbox}
\end{subtable}
\label{tab:simulation_lambda_2}
\end{table}


\section{Application}
\label{sec:application}
\begin{figure}[ht]
    \centering
        \includegraphics[width=\linewidth]{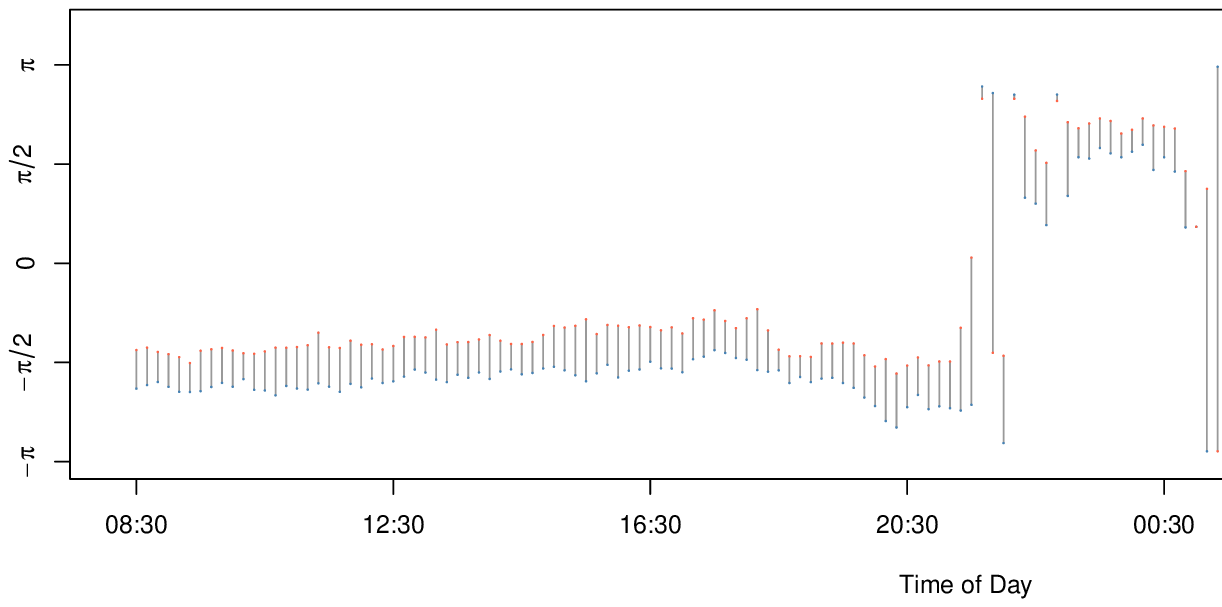}
        \includegraphics[width=\linewidth]{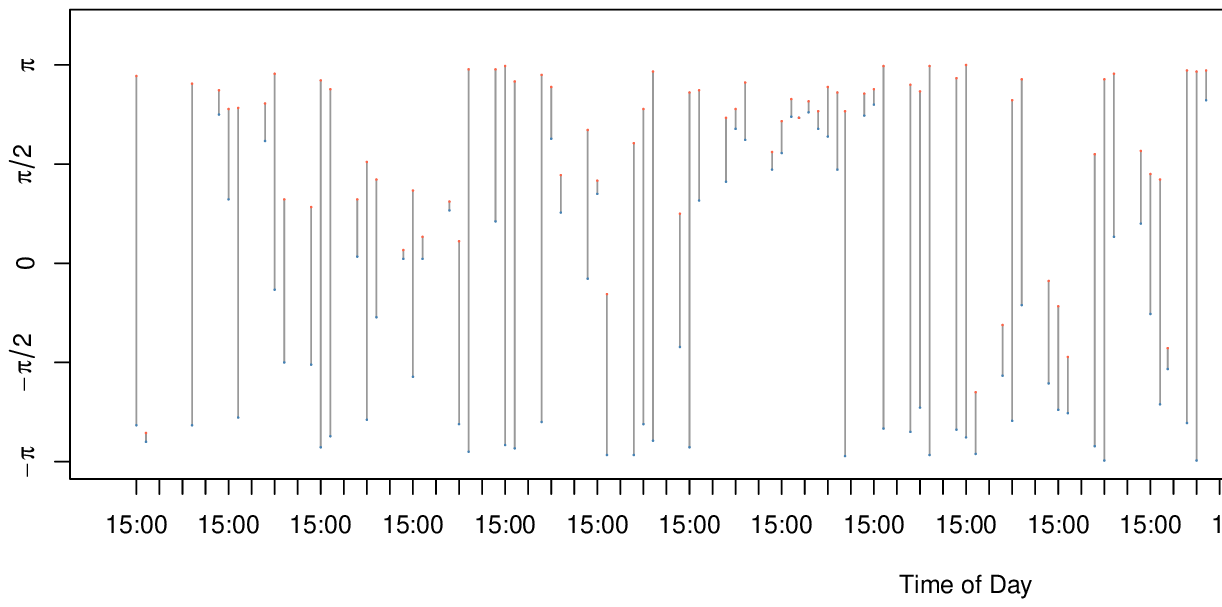}
    \caption{Interval-valued wind direction data for South Africa (top) and Brazil (bottom).}
    \label{fig:applications}
\end{figure}

The practical utility of the proposed model is illustrated using interval-valued wind direction time series data recorded at two locations. The first being a South African wind farm where data was recorded every 10 minutes, and the second being a Brazilian meteorological station where the data was recorded every hour and summarised into 5-hourly intervals to create a wide arc. 
\noindent In Figure \ref{fig:applications}, the interval-valued wind directions are plotted for the two locations. In the first plot, it can be seen that the data is situated on narrow arcs away. Conversely, in the second plot, the data is spread widely across the circle near the circular cut. Thus, the proposed approach is evaluated for different scenarios that could occur naturally in the real domain.

The proposed IN-DCS model is compared to the CINAR model of \cite{mohan2026cinar}.
We consider an interval-valued performance metric, the circular mean-squared error, IN-CMSE, which is defined as
\begin{equation}
\text{IN-CMSE}(\widehat{\Theta},\Theta) = \frac{1}{T-p}\sum_{t=1}^{T}\left[d\!\left(\theta_t^{L},\widehat{\theta}_t^{L}\right) + d\!\left(\theta_t^{U},\widehat{\theta}_t^{U}\right)\right], ~ d(\theta,\widehat{\theta}) = 1 - \cos(\theta - \widehat{\theta}),
\label{eq:incmse}
\end{equation}
where $p, T$ are the number of parameters and observations, respectively. To test  $\lambda= 0$, we make use of a likelihood ratio test (LRT) \citep{casella2024statistical}. The test statistic is defined as $\Lambda= -2(\ell_1-\ell_2)$, where $\ell_1$ is the log-likelihood of the independent IN-DCS model, when $\lambda=0$ whilst $\ell_2$ is the log-likelihood of the dependent IN-DCS model. The test statistic is chi-square distributed with degrees of freedom equal to the difference in the number of parameters estimated under the two models, in this case $df=1$. We make use of 5\% level of significance.

Table~\ref{tab:app1} and Table~\ref{tab:dcs_application2} report the estimated parameters of the proposed model for both datasets, log-likelihood, Bayesian Akaike Information Criteria (BIC), IN-CMSE and the LRT, respectively.  From Table \ref{tab:app1}, the dependent IN-DCS model provides a better fit than the independent alternative. Furthermore, the LRT leads to rejection of the null hypothesis $H_0:\lambda = 0$. These findings suggest that the two interval endpoints evolve jointly and that including a dependence parameter improves model performance for highly concentrated wind-direction settings. In contrast, the Brazil dataset, in Table \ref{tab:dcs_application2}, produces very different results. The LRT of independence is not rejected. Consistent with this result, the estimated dependence parameter is close to zero. In this case, the simpler independent IN-DCS model is preferred for the observed interval-valued data.

\begin{figure}[ht]
    \centering
        \includegraphics[width=0.8\linewidth]{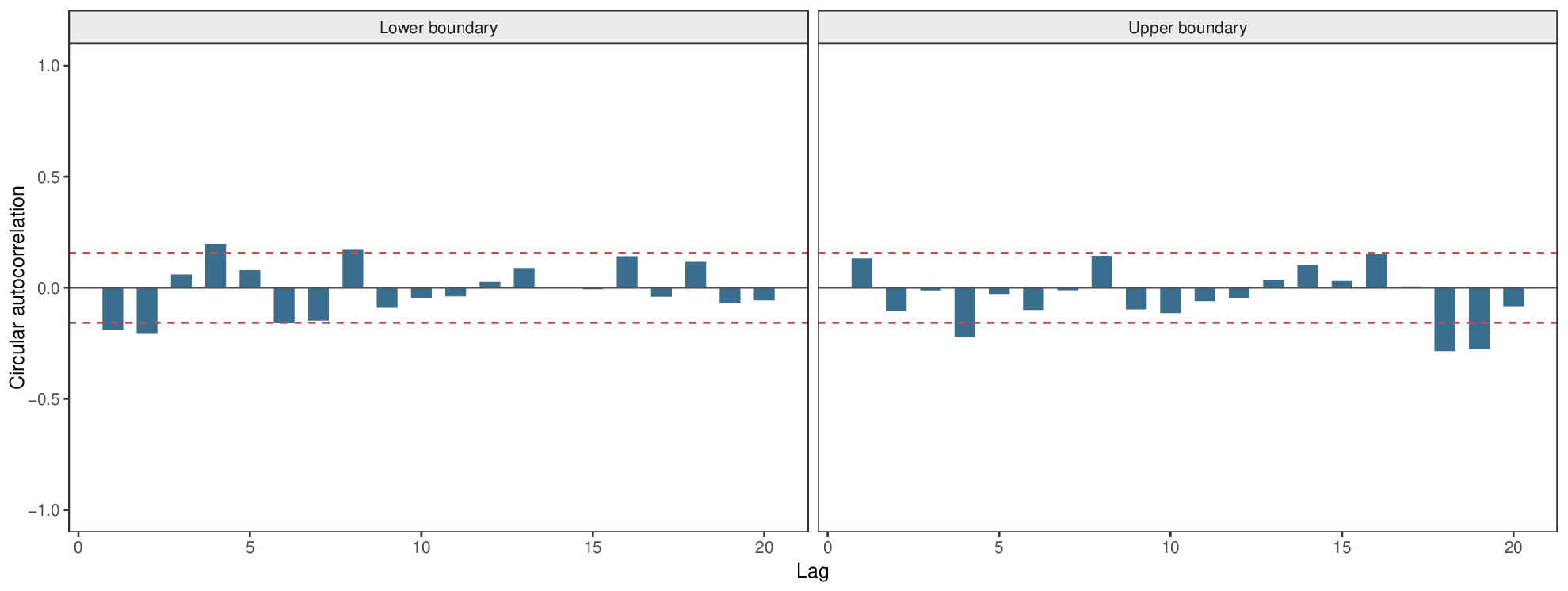}
        \includegraphics[width=0.8\linewidth]{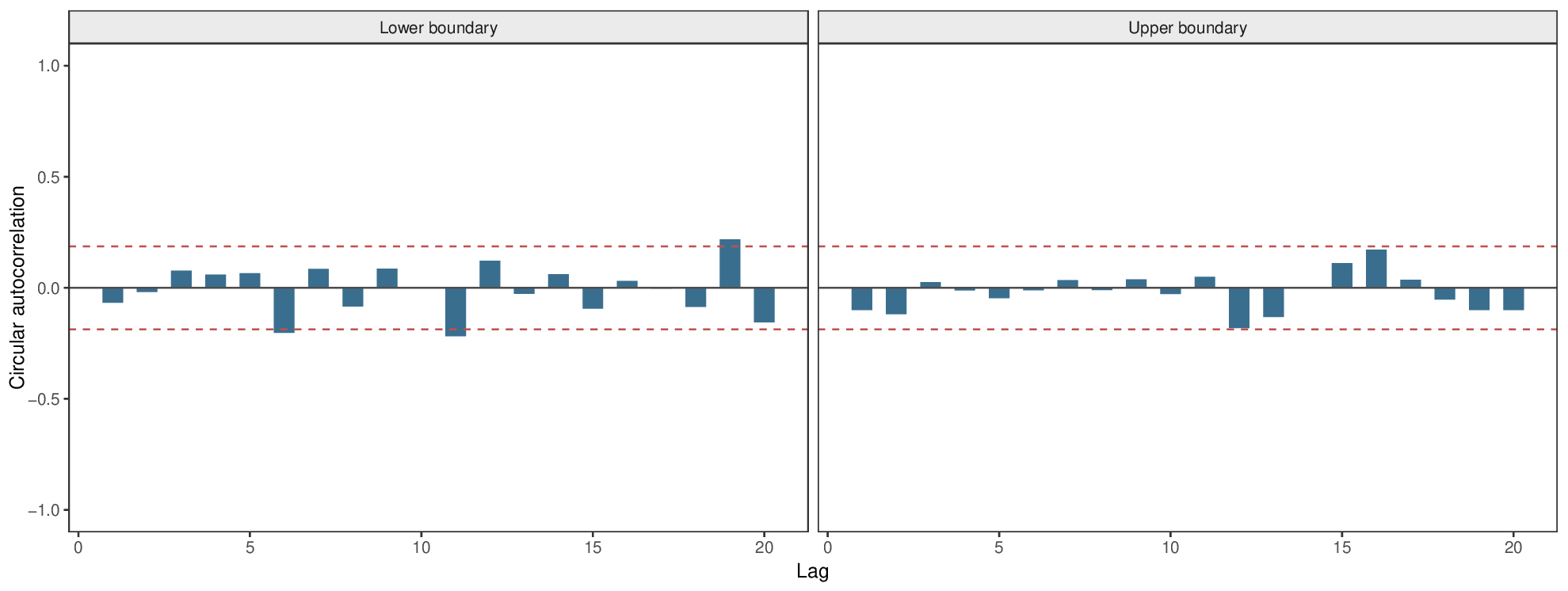}
    \caption{CACF of the differenced observations for South Africa (top) and Brazil (bottom).}
    \label{fig:CACF}
\end{figure}

Since the time series coefficients ($\phi$) for the South African data are near the boundary, the sample circular Autocorrelation functions (CACF) as given in \cite{harvey2024modelling}, of the differenced lower and upper boundaries, are plotted in Figure \ref{fig:CACF}. This plot indicates that the persistence in the boundaries is captured, and no additional dynamic structure is required. Even though the estimated time series coefficients for Brazil are not near the boundary, as an additional check, the CACF is also plotted in Figure \ref{fig:CACF} and it is as expected. 

For comparison, the CINAR of \cite{mohan2026cinar} was fitted to both datasets for orders $p = 1, \dots, 5$. Since the CINAR likelihood is defined using the internal distribution of the intervals, two models are compared using the IN-CMSE since the BICs are not comparable. The results for both datasets are given in Tables~\ref{tab:incme_cinar} and~\ref{tab:incme_cinar_brazil}.  
For the South African data, the smallest CINAR IN-CMSE in Table \ref{tab:CINAR_SA} is obtained at $p = 1$. The results of the IN-CMSE confirm that the IN-DCS model outperforms the CINAR model. 
For the Brazilian data, IN-CMSE of the CINAR model in Table \ref{tab:CINAR_BRA} decreases with $p$. This highlights the CINAR model's limitation to a narrow arc away from the circular cut. For the Brazil dataset, the results of the IN-CMSE also confirm the good predictive performance of the proposed IN-DCS model and the suitability of the fully rotational invariant IN-DCS model in the presence of wide arcs near the circular boundaries. 
\begin{table}[ht]
\centering
\begin{threeparttable}
\caption{Estimated parameters of the IN-DCS model fitted to the wind dataset
in South Africa, $T = 156$.}
\label{tab:dcs_sa}
\begin{tabular}{l *{4}{S[table-format=-1.4]}}
\toprule
 & \multicolumn{2}{c}{ Independent} & \multicolumn{2}{c}{\bf Dependent} \\
\cmidrule(lr){2-3} \cmidrule(lr){4-5}
 & {$L$} & {$U$} & {$L$} & {$U$} \\
\midrule
$\hat\kappa$   & 1.3220  & 1.3754  & 1.0702  & 1.1921  \\
$\hat\phi$     & 0.9542  & 0.9008  & 0.9981  & 0.9426  \\
$\hat\mu$      & -2.2113 & -2.3420 & -2.2012 & -2.0962 \\
$\hat\upsilon$ & 5.9506  & 6.2189  & 6.4341  & 6.6115  \\
$\hat\lambda$  & {--}    & {--}    & \multicolumn{2}{c}{1.7537} \\
\midrule
$\log L$ & \multicolumn{2}{c}{$-176.038$} & \multicolumn{2}{c}{$-167.862$} \\
BIC      & \multicolumn{2}{c}{392.476}    & \multicolumn{2}{c}{\bf 381.174} \\
IN-CMSE   & \multicolumn{2}{c}{0.1728}     & \multicolumn{2}{c}{\bf 0.1726} \\
\bottomrule
\end{tabular}\label{tab:app1}
\begin{tablenotes}\footnotesize
\item Likelihood-ratio test of $H_0\!: \lambda = 0$: $LR = 16.352$, the
critical value $\chi^2_{1,\,0.05} = 3.841$ ($p < 0.001$). The best model based on LRT, BIC and IN-CMSE is bolded.
\end{tablenotes}
\end{threeparttable} 
\end{table}

\begin{table}[ht]\centering
\begin{threeparttable}
\caption{IN-CMSE of the CINAR model for orders $p = 1, \dots, 5$ fitted to the wind direction dataset in South Africa.}
\label{tab:incme_cinar}
\begin{tabular}{cccccc}
\toprule
$p$ & Uniform & Triangular & Pert & U-quadratic  \\
\midrule
1 & \textbf{0.2263} & \textbf{0.2263} & \textbf{0.2263} & { 0.2264} \\
2 & 0.2407 & 0.2407 & 0.2407 & 0.2406  \\
3 & 0.2271 & 0.2271 & 0.2271 & 0.2270  \\
4 & 0.2427 & 0.2429 & 0.2429 & 0.2425  \\
5 & 0.2423 & 0.2424 & 0.2424 & 0.2422  \\
\bottomrule
\end{tabular} \label{tab:CINAR_SA}
\begin{tablenotes}\footnotesize
\item Smallest value is bolded.
\end{tablenotes}
\end{threeparttable}
\end{table}

\begin{table}[H]
\centering
\begin{threeparttable}
\caption{Estimated parameters of the IN-DCS model fitted to the wind dataset
in Brazil, $T = 111$.}
\label{tab:dcs_application2}
\begin{tabular}{l *{4}{S[table-format=-1.4]}}
\toprule
 & \multicolumn{2}{c}{\bf Independent} & \multicolumn{2}{c}{ Dependent} \\
\cmidrule(lr){2-3} \cmidrule(lr){4-5}
 & {$L$} & {$U$} & {$L$} & {$U$} \\
\midrule
$\hat\kappa$   & 0.5153  & 0.6057  & 0.4907  & 0.5578  \\
$\hat\phi$     & 0.7289  & -0.0493 & 0.7396  & -0.1085 \\
$\hat\mu$      & -2.8104 & 2.3602  & -2.8125 & 2.3146  \\
$\hat\upsilon$ & 1.0386  & 1.4902  & 1.0483  & 1.4976  \\
$\hat\lambda$  & {--}    & {--}    & \multicolumn{2}{c}{$-0.2901$} \\
\midrule
$\log L$ & \multicolumn{2}{c}{$-339.630$} & \multicolumn{2}{c}{$-338.609$} \\
BIC      & \multicolumn{2}{c}{\bf 716.937}    & \multicolumn{2}{c}{719.603} \\
IN-CMSE   & \multicolumn{2}{c}{\bf 0.9464}     & \multicolumn{2}{c}{ 0.9532} \\
\bottomrule
\end{tabular}
\begin{tablenotes}\footnotesize
\item Likelihood-ratio test of $H_0\!: \lambda = 0$: $LR = 2.044$,
critical value $\chi^2_{1,\,0.05} = 3.841$ ($p = 0.153$). The best model based on LRT, BIC and IN-CMSE is bolded.
\end{tablenotes}
\end{threeparttable}
\end{table}

\begin{table}[H]
\centering
\begin{threeparttable}
\caption{IN-CMSE of the CINAR model for orders $p = 1, \dots, 5$, fitted to the wind direction dataset in Brazil.}
\label{tab:incme_cinar_brazil}
\begin{tabular}{ccccc}
\toprule
$p$ & Uniform & Triangular & Pert & U-quadratic \\
\midrule
1 & 2.7271 & 2.7271 & 2.7271 & 2.7272  \\
2 & 2.6948 & 2.6947 & 2.6946 & 2.6950 \\
3 & 2.6007 & 2.6005 & 2.6004 & 2.6011  \\
4 & 2.5485 & 2.5481 & 2.5480 & 2.5492  \\
5 & {2.4906} & {2.4900} & \textbf{2.4899} & {2.4915} \\
\bottomrule
\end{tabular}
\label{tab:CINAR_BRA}
\begin{tablenotes}\footnotesize
\item Smallest value is bolded.
\end{tablenotes}
\end{threeparttable}
\end{table}
In conclusion, the application study demonstrates the flexibility of the proposed IN-DCS framework across two interval-valued wind direction datasets with very different characteristics. These results support the use of a rotationally invariant, score-driven approach for interval-valued circular time series.
\section{Concluding Remarks}
\label{sec:conclusion}
This paper proposed a score-driven framework for interval-valued circular time series data. Unlike the existing approach of \cite{mohan2026cinar} that relies on a transformation to the real line, the proposed IN-DCS framework models lower and upper interval boundaries directly and jointly on the torus, thereby preserving both the interval structure of the observations and the inherent periodicity of the circular domain. The methodology enables dependence between interval boundaries through a toroidal distribution and a jointly derived score function. The simulation study provided evidence for the finite-sample performance of the proposed estimation procedure. Across a wide range of parameter configurations, including independent and dependent boundary structures, the estimators exhibited numerical stability, robustness and consistency of the maximum likelihood estimators. The practical utility of the methodology was demonstrated using interval-valued wind direction data from South Africa and Brazil.  Furthermore, the likelihood ratio test showed that dependence between interval boundaries can be important in practice, as observed in the South African dataset, while in other settings, such as the Brazilian dataset, a simpler independent specification may be adequate. Hence, the methodology provides a new approach to circular interval-valued time series and thus extends the applicability of score-driven methods to the field of toroidal data. Future research may consider extending the framework to incorporate higher-order dependence structures and multivariate interval-valued circular responses for interval-valued directional processes.

\section* {Funding} 
{\small The work of Najmeh Nakhaei Rad is supported in part by the National Research
Foundation (NRF) of South Africa, Ref.: RA210106581084, grant No. 150170, and the UCDP grant at the University of Pretoria, South Africa. The opinions expressed and conclusions arrived at are those of the
authors and are not necessarily to be attributed to the funders.}

\section*{Conflict of Interest}

The authors declare that they have no conflicts of interest.

\section*{Data Availability}
The Brazilian data used in this study are publicly available, and the South African data will be available on request from the corresponding author. 

\bibliographystyle{apalike}
\bibliography{references}

\end{document}